\documentclass[iop]{emulateapj}
\usepackage{natbib}
\usepackage{amsmath,bm}
\usepackage{graphicx}
\usepackage{enumerate}
\usepackage{multirow}
 \usepackage{times}
 \usepackage{mathptmx}     
\usepackage{comment}

\usepackage{color}           

\usepackage{url}
\usepackage[colorlinks, citecolor=blue]{hyperref}

\newcommand{\Pa}{\Phi_\mathrm{axi}}
\newcommand{\Fp}{F_\mathrm{pol}}

\begin{document}

\title{Magnetohydrodynamic Modeling of the Solar Eruption on 2010
       April 8}

\author{B. Kliem\altaffilmark{1,2,3,4}, 
        Y. N. Su\altaffilmark{5}, 
        A. A. van Ballegooijen\altaffilmark{5}, and
        E. E. DeLuca\altaffilmark{5}}

\affil{$^1$Institute of Physics and Astronomy, University of Potsdam,
           Potsdam 14476, Germany}
\affil{$^2$Mullard Space Science Laboratory, University College London,
           Holmbury St.\ Mary, Dorking, Surrey, RH5 6NT, UK}
\affil{$^3$Yunnan Astronomical Observatory, Chinese Academy of
           Sciences, Kunming 650011, China}
\affil{$^4$College of Science, George Mason University, Fairfax, VA
           22030, USA}
\affil{$^5$Harvard-Smithsonian Center for Astrophysics, Cambridge, MA
           02138, USA}

\shorttitle{Solar Eruption on 2010 April 8}  
\shortauthors{Kliem et al.}

\email{bkliem@uni-potdam.de}

\journalinfo{Manuscript ms\_8apr, revised 2013 October 4} 
\submitted{Received 2013 May 24; accepted 2013 October 4; published
           2013 December 3}

\begin{abstract}

\noindent The structure of the coronal magnetic field prior to
eruptive processes and the conditions for the onset of eruption are
important issues that can be addressed through studying the
magnetohydrodynamic (MHD) stability and evolution of nonlinear force-free
field (NLFFF) models. This paper uses data-constrained NLFFF models of
a solar active region (AR) that erupted on 2010 April~8 as initial
condition in MHD simulations. These models, constructed with the
techniques of flux rope insertion and magnetofrictional relaxation,
include a stable, an approximately marginally stable, and an unstable
configuration. The simulations confirm previous related results of
magnetofrictional relaxation runs, in particular that stable flux rope
equilibria represent key features of the observed pre-eruption coronal
structure very well and that there is a limiting value of the axial
flux in the rope for the existence of stable NLFFF equilibria. The
specific limiting value is located within a tighter range, due to the
sharper discrimination between stability and instability by the MHD
description. The MHD treatment of the eruptive configuration yields
very good agreement with a number of observed features like the
strongly inclined initial rise path and the close temporal association
between the coronal mass ejection and the onset of flare reconnection.
Minor differences occur in the velocity of flare ribbon expansion and
in the further evolution of the inclination; these can be eliminated
through refined simulations. We suggest that the slingshot effect of
horizontally bent flux in the source region of eruptions can
contribute significantly to the inclination of the rise direction.
Finally, we demonstrate that the onset criterion formulated in terms
of a threshold value for the axial flux in the rope corresponds very
well to the threshold of the torus instability in the considered AR.

\end{abstract}


\section{Introduction}
\label{s:introduction}

The coronal source regions of solar eruptions (eruptive prominences,
coronal mass ejections (CMEs), and flares) are characterized by low plasma
beta, especially in active regions (ARs), where $\beta\sim10^{-3}$
\citep{Gary2001}. Consequently, the magnetic field must remain nearly
force free in the equilibrium sequence seen as the quasi-static
evolution toward the onset of eruption. Since the energy needed to
power the eruption can only be stored in coronal currents
\citep{Forbes2000} and since these currents tend to be strongly
concentrated in the core of the source region
\cite[e.g.,][]{XSun&al2012}, the pre-eruptive field is, to a good
approximation, a nonlinear force-free field (NLFFF) obeying
$\bm{\nabla}\bm{\times}\bm{B}=\alpha\bm{B}$ with the ``force-free parameter''
$\alpha(\bm{r})$ being a scalar function that varies across $\bm{B}$.

In order to understand the mechanism of eruptions and to quantify the
criteria for their onset, knowledge of the field in the erupting
coronal volume is required. However, the three-dimensional (3D) field
structure is not amenable to measurement, and reliably inferring the
coronal NLFFF from photospheric magnetograms has proven very
difficult, even if vector magnetograms are available. Two strategies
have been pursued. \emph{Extrapolation} techniques solve the
force-free equation numerically using a vector magnetogram as the bottom
boundary condition. Several schemes produced reliable results when
applied to test fields, in particular but not exclusively, the
Grad-Rubin iteration, an optimization scheme, and magnetofrictional
relaxation (MFR); see \citet{Schrijver&al2006} for an overview. However, it
appears that the reconstruction of the coronal field from observed
vector magnetograms matches reality closely only in a still relatively
small number of cases; see especially \citet{Schrijver&al2008a},
\citet{Canou&Amari2010}, \citet{YellesChaouche&al2012},
\citet{Valori&al2012}, and \citet{XSun&al2012}. The methods had
difficulty in producing reliable results for some other cases
\cite[e.g.,][]{Metcalf&al2008, Schrijver&al2008a, DeRosa&al2009}. The
violation of the force-free condition at the photospheric level in
part of the magnetogram, uncertainties of the transverse magnetogram
components, and the fragmentation of the flux in the photosphere down
to sub-resolution scales are supposed to cause the problems which are
not yet clearly understood.

The extrapolation technique has supported the modeling of eruptions as
a flux rope instability \citep{Torok&Kliem2005, Kliem&Torok2006} by
computing stable and unstable coronal fields containing a flux
rope \citep{Valori&al2010} from magnetograms of the active-region model by
\citet{Titov&Demoulin1999}. However, to date only few extrapolations
of observed magnetograms have found a flux rope, see, e.g.,
\citet{Y.Yan&al2001}, \citet{Canou&Amari2010}, and
\citet{YellesChaouche&al2012}. A counterexample is the reconstruction
of the highly nonpotential AR NOAA 11158, which did not
yield a flux rope
(\citeauthor{Nindos&al2012} \citeyear{Nindos&al2012}; G.~Valori 2013, private
communication), although the subsequent violent eruption was
suggestive of a flux rope eruption \citep{Schrijver&al2011,
Zharkov&al2011}.

The \emph{flux rope insertion} method \citep{vanBallegooijen2004,
vanBallegooijen&al2007} provides a viable alternative tool for the
determination of the coronal NLFFF. Here a flux rope is inserted in
the potential field computed from the vertical magnetogram component,
to run at low height along the magnetogram's polarity inversion line
(PIL) in the section where a filament channel indicates a nearly
horizontal, current-carrying field. This configuration is then
numerically relaxed using magnetofriction \citep{Yang&al1986}. The
resulting numerical equilibrium has received observational support in
a growing number of cases by favorable comparison of various field
line shapes (dips, arches of various shear angle relative to the PIL,
S shape) and quasi-separatrix layers with coronal structures
\citep{Bobra&al2008, Su&al2009a, Su&al2011, Su&vanBallegooijen2012,
Savcheva&vanBallegooijen2009, Savcheva&al2012a, Savcheva&al2012b,
Savcheva&al2012c}. In these studies, quiet-Sun or decaying active
regions were modeled, where persistent shearing and flux cancellation
make the formation of a coronal flux rope likely. It should be noted,
however, that the method is not restricted to configurations
containing a flux rope; the numerical relaxation can also result in an
arcade configuration \citep{Su&al2011}. 

The application of the flux
rope insertion method led to the suggestion that the ratio between the
magnetic flux in the rope, primarily the axial flux, and the unsigned
flux in the AR may provide a new onset criterion for
eruptions \citep{Bobra&al2008}. This has found considerable support in
subsequent applications \cite[see in particular][]{Su&al2011} as well
as in magnetohydrodynamic (MHD) simulations of flux cancellation
\citep{Aulanier&al2010,
Amari&al2010}, although the threshold may vary in a wide range
\citep{Savcheva&al2012c}. 

While the MFR
method is well suited to find the dividing line between stability and
instability in parameter space, it fails in correctly describing the
dynamic evolution of unstable configurations because it represents a
strongly reduced MHD description which disregards the inertia of
the plasma and thus excludes the MHD waves. It is
therefore of interest to study the relaxed or partially relaxed
configurations in full MHD simulations. This allows further judgment
of the obtained configurations by comparison with observations of
erupting regions, and, additionally, an independent test of the
marginal stability line in parameter space. The present investigation
aims at carrying out the first such experiments. 

\citet[][hereafter Paper~I]{Su&al2011} performed extensive and
detailed modeling of NOAA AR 11060 at the stage just
prior to its eruption on 2010 April~8. The event included an erupting
filament that evolved into a coronal mass ejection (CME) accompanied
by a moderate flare. By varying the parameters of the inserted flux
rope in a range of axial ($\Pa$) and poloidal ($\Fp$) flux, the
stability boundary in the $\Pa$--$\Fp$ plane was determined with
relatively high accuracy. Weakly twisted and arching field lines of a
resulting stable configuration rather close to the stability boundary
showed good correspondence to X-ray and EUV loops prior to the
eruption. Low-lying highly sheared field lines of an unstable
configuration rather close to the stability boundary corresponded well to the
observed initial flare loops. A further result of high interest, with
regard to the onset of eruptions, was that an X-type magnetic structure
known as a hyperbolic flux tube \cite[HFT;][]{Titov&al2002, Titov2007}
appeared at low heights under the flux rope when the axial flux was
raised to approach the stability boundary. In the following we study a
series of configurations from this investigation in the zero-beta MHD
approximation. The configurations are arranged on a path in the
$\Pa$--$\Fp$ plane that crosses the stability boundary and include the
models which compare favorably to the observations.

The simulations also allow us to compare the conditions at the onset
of instability with the ratio of rope and ambient flux
\citep{Bobra&al2008} and with the height profile of the flux rope's ambient
field \citep{vanTend&Kuperus1978, Kliem&Torok2006}. Identification of
the conditions necessary for instability is a key goal of this
research.

Section~\ref{s:observations} presents a summary of the observations
from \citet{Su&al2011}. This is followed by a detailed description of
the numerical model in Sections~\ref{s:MF} and \ref{s:MHD}. The
relaxation of stable configurations and the eruption of an unstable
configuration are presented in Sections~\ref{s:relaxation} and
\ref{s:eruption}, respectively. Our conclusions are
given in Section~\ref{s:conclusions}.

\section{Summary of Observations}
\label{s:observations}

\begin{figure*}[t]                                               
\centering
\includegraphics[width=.64\linewidth]{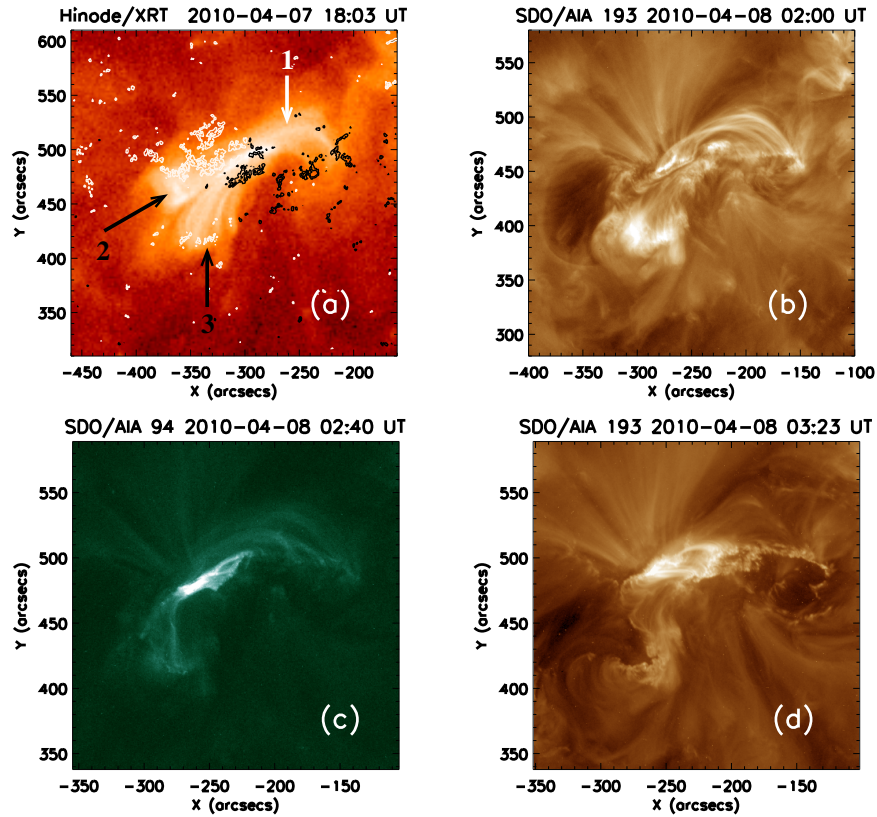}
\caption{\label{fig1}
 Structure of AR~11060 before and during the eruption observed by
 (a) \textsl{Hinode}/XRT, and (b)--(d) \textsl{SDO}/AIA.
 In panel~(a) the contours refer to the photospheric line of sight
 (LOS) magnetogram taken by the Helioseismic and Magnetic Imager
 \cite[HMI;][]{Schou&al2012} on \textsl{SDO} at 02:00~UT on 2010
 April~8 (white: positive; black: negative; see Figure~\ref{fig2}
 for a more detailed contour plot). The coronal loops in the
 core of the region are organized in three sets, labeled as 1--3 and
 discussed in the text.
 (A color version of this figure is available in the online journal.)} 
\end{figure*}

The event under study originated in AR~11060 on 2010 April~8 around
02:30~UT. It included an eruptive prominence that evolved into a CME,
a two-ribbon solar flare of \textsl{GOES} class B3.7, prominent coronal
dimmings, and a coronal EUV wave. The eruption was observed on disk by
the Atmospheric Imaging Assembly \cite[AIA,][]{Lemen&al2012} on the
\textsl{Solar Dynamics Observatory (SDO)} and by the X-Ray Telescope
\cite[XRT,][]{Golub&al2007} on the \textsl{Hinode} mission. The
Extreme Ultraviolet Imager (EUVI) and the coronagraphs COR1 and COR2
\citep{Howard&al2008} aboard the \textsl{Solar-Terrestrial Relations
Observatory (STEREO) Ahead} spacecraft saw the event nearly  exactly
at the east limb. A detailed description of these data including
animations can be found in Paper~I. Additional information is given in
the analysis of the EUV wave \citep{W.Liu&al2010} and of the possible
occurrence of the Kelvin-Helmholtz instability at the surface of the
expanding flux \citep{Ofman&Thompson2011}.

Figure~\ref{fig1} shows \textsl{Hinode}/XRT (a) and \textsl{SDO}/AIA
(b) images of AR~11060 before the eruption. The core of this active
region contains three sets of coronal loops as shown in
Figure~\ref{fig1}(a). The set of loops labeled 1 is located in the
northwestern part of the region (white arrow). Loop sets~2 and 3 are
associated, respectively, with the major positive polarity (white
contours) in the northeast of the region and with a group of minor
positive flux patches in the southeast (both marked by black arrows).
The combined loop sets~1 and 2 yield a slightly sigmoidal appearance
at soft X-rays. Figure~\ref{fig1}(b) shows a thin, dark filament
following mainly the highly sheared loop set~2, i.e., located between
loop sets~1 and 3.

The eruption began around 02:10~UT with motion of material along this
filament in the southeastward direction (as seen by AIA) and nearly
horizontally (as seen by EUVI-A on \textsl{STEREO Ahead}). About
18~minutes later, the filament began to lift off. From about 02:33~UT,
the overlying loops are seen to rise in the EUVI-A 195~{\AA} images.
The first flare brightenings appear in AIA 94~{\AA} images on either
side of the erupting filament around 02:30~UT, and the \textsl{GOES}
soft X-ray emissions commence near 02:45~UT. Thus, the onset of upward
flux expansion, which evolved into the CME, and the onset of the flare
brightenings occurred in close temporal and spatial association.

Initially highly sheared flare loops became visible between 02:30 and
02:40~UT in the AIA 94~{\AA} channel (Figure~\ref{fig1}(c)), they
connected the first flare brightenings in the strong-field section of
the PIL. The typical evolution toward lower shear was seen in the
loops forming subsequently (Figure~\ref{fig1}(d)). These observations
indicate that the component of the magnetic field along the PIL points
northwestward and that the field has positive helicity, consistent
with the forward S shape of the combined loop sets~1 and 2. The
brightenings early in the flare (Figure~\ref{fig1}(c)) as well as the
subsequent equatorward growth of the CME dimmings yield an overall
inverted U shape of the parts of the region actively involved in the
eruption. This is similar to the combined loop sets~1 and 3 and
suggests that their flux is an integral part of the eruption, although
loop set~3 connects to a minor flux area in the magnetogram.

The erupting flux took a strongly inclined path toward the equator,
initially at about $45^\circ$ from vertical, then developing an even
more inclined southward expansion in the low corona imaged by EUVI-A,
and finally turning into a more radial propagation near the
equatorial plane, as imaged by the COR2-A coronagraph. The EUVI data
do not reveal the rise profile of the erupting filament, as the
material very quickly lost contrast in the 195~{\AA} channel and the
cadence of the other channels was too low. However, the rise velocity
of overlying coronal loops can be estimated and is found to reach
about 170~km~s$^{-1}$ within the EUVI-A field of view (see
Section~\ref{ss:scaling}). The CME reached a median projected speed of
$\approx520$~km~s$^{-1}$ in the COR2-A field of view, as quoted in the
CACTus CME Catalog\footnote{\href{http://secchi.nrl.navy.mil/cactus/}{http://secchi.nrl.navy.mil/cactus/}}
\citep{Robbrecht&al2009}.

\section{Magnetofrictional Modeling}
\label{s:MF}

The presence of highly sheared loops (Figure~\ref{fig1}) indicates
that the coronal magnetic field in the observed region deviates
significantly from a potential field. In Paper~I we used
a set of codes, called the Coronal Modeling System (CMS), to construct
non-potential field models of the observed region. The methodology
involves inserting a thin magnetic flux rope into a three-dimensional
(3D) potential-field model along a specified path, and then applying
magnetofrictional relaxation (MFR) \citep{Yang&al1986,
vanBallegooijen&al2000} to produce either a non-linear force-free
field (NLFFF) model or unstable model, depending on the magnitude of
the inserted flux. MFR refers to an evolution of the magnetic field
$\bm{B} (\bm{r},t)$ in which the medium is assumed to be highly
conducting and the plasma velocity $\bm{v}$ is proportional to the
Lorentz force $\bm{j \times B}$, where $\bm{j = \nabla \times B}$ is
the current density. The resulting 3D magnetic models are based on
observed photospheric magnetograms and therefore accurately represent
the lower boundary conditions on the magnetic field in the flaring
active region. This makes such models ideally suited as initial
conditions for 3D MHD simulations.

In this paper, we use flux rope insertion and MFR to produce initial
conditions for a 3D MHD code, which is described in
Section~\ref{s:MHD}. One problem is that the MHD code assumes
Cartesian geometry, whereas the CMS codes use spherical geometry,
and the considered area is too large to ignore curvature effects.
Therefore, we cannot simply interpolate the relevant models from
Paper~I. Instead, the methods used in Paper~I are modified in several
ways, as described in the following. First, we construct a map of
$B_r (\theta,\phi)$, the radial component of magnetic field at the
solar surface ($r=R_\odot$) in and around the observed AR.
The map is based on an \textsl{SDO}/HMI magnetogram taken at 2:00~UT.
The map is computed by interpolating the observed LOS field onto a
longitude-latitude grid, and then dividing by the cosine of the
heliocentric angle to obtain $B_r$. This original map has
$384 \times 384$ cells and covers an area of $41.3^\circ$ in longitude
by $36.8^\circ$ in latitude, centered on the observed AR.
Then the spatial scale of the map is reduced, and the center of the
map is displaced to the equator, corresponding to a reduction in scale
by a factor $f = 16.97$. The rescaled map covers only $2.2^\circ$ in
longitude and latitude, which is sufficiently small that the resulting
spherical models can be treated as if they were Cartesian.

\begin{figure}[t]                                                
\centering
\includegraphics[width=\linewidth]{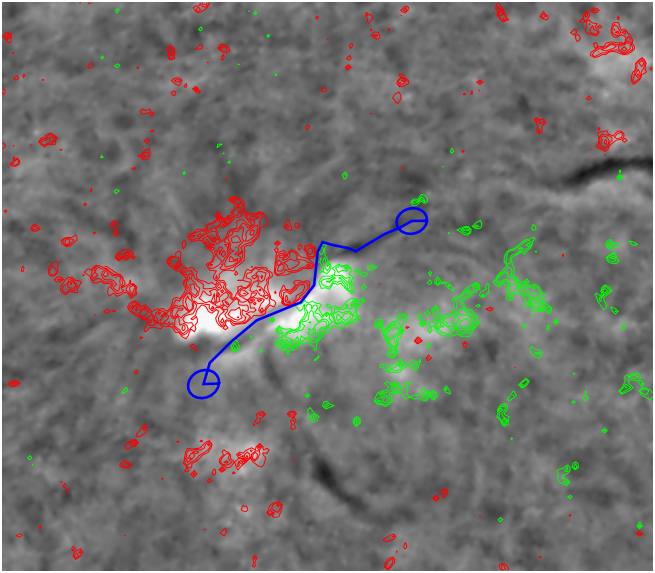}
\caption{\label{fig2}
 H$\alpha$ image of the observed active region, obtained at
 Kanzelh\"{o}he Solar Observatory on 2010 April~8 at 9:00~UT. The red
 and green contours show the radial photospheric magnetic field
 component derived
 from the \textsl{SDO}/HMI LOS magnetogram taken at 2:00~UT. The blue
 curve shows the path of the inserted flux rope between the marked
 end points.
 (A color version of this figure is available in the online journal.)} 
\end{figure}

In CMS the magnetic field is described using two spatial grids: a high
resolution grid (HIRES) covering the target region and its local
surroundings, and a low resolution grid covering the entire Sun. The
non-potential field in the HIRES region is described in terms of
vector potentials ($\bm{B = \nabla \times A}$). In the standard
approach, the global grid is used for a potential field source surface
(PFSS) model; its main purpose is to improve the side boundary
conditions for the HIRES domain. However, for the scaled models used
here the global grid can no longer be used. Instead we assume that the
normal component of magnetic field vanishes at the side boundaries of
the HIRES domain ($B_\theta = 0$ and $B_\phi = 0$ at the latitudinal
and longitudinal boundaries, respectively). These boundary conditions
imply that the net flux entering the domain through the lower boundary
is also present at larger heights, and leaves the domain through the
outer boundary.  We found that small flux imbalances between positive
and negative magnetic fluxes in the imposed magnetic map ($B_r$) can
produce fields at large heights that are dominated by such ``monopole''
components. Furthermore, in preliminary MHD simulations we found that
such monopolar fields can inhibit the eruption of low-lying flux
ropes. Therefore, in the present paper we correct the flux imbalance
in $B_r (\theta,\phi)$ at the lower boundary. The photospheric fluxes
are balanced by subtracting a spatially constant value of about
2.2~Gauss from $B_r (\theta,\phi)$, except near the border of the
computational domain where $B_r$ is set to zero.

The next steps are to select the path of the flux rope, compute a
potential field, and insert the flux rope into the 3D magnetic model.
The path should follow the PIL of the AR;
we use the same path as in Paper~I. Figure~\ref{fig2} shows
the selected path superposed on a H$\alpha$ image obtained at
Kanzelh\"{o}he Solar Observatory at 9:00~UT. The red and green
contours indicate the photospheric flux distribution
$B_r (\theta,\phi)$ before scaling.

In this paper, we construct three different models with the same flux
rope path, height, and cross section, but with different values of the
axial flux of the flux rope. The axial fluxes before scaling are
$\Pa = 4 \times 10^{20}$, $5 \times 10^{20}$ and
$6 \times 10^{20}$~Mx, which were chosen because these values appear
to straddle the stability boundary (see Paper~I). After scaling these
axial fluxes are reduced by the square of the scaling factor. The
axial flux is inserted into the 3D model as a thin flux tube that is
elevated above the photosphere. At the ends of the tube (indicated by
circles in Figure~\ref{fig2}), the flux of the tube is connected to
surrounding sources on the photosphere. Poloidal fields are added by
wrapping flux rings around this tube. The assumed poloidal flux is
$\Fp=10^{10}$~Mx~cm$^{-1}$ (before reduction by the scaling factor).
For more details on how the flux rope is inserted into the 3D models,
see \citet{Su&al2011} and references therein.

\begin{figure*}[t]                                               
\centering
\includegraphics[width=.82\linewidth]{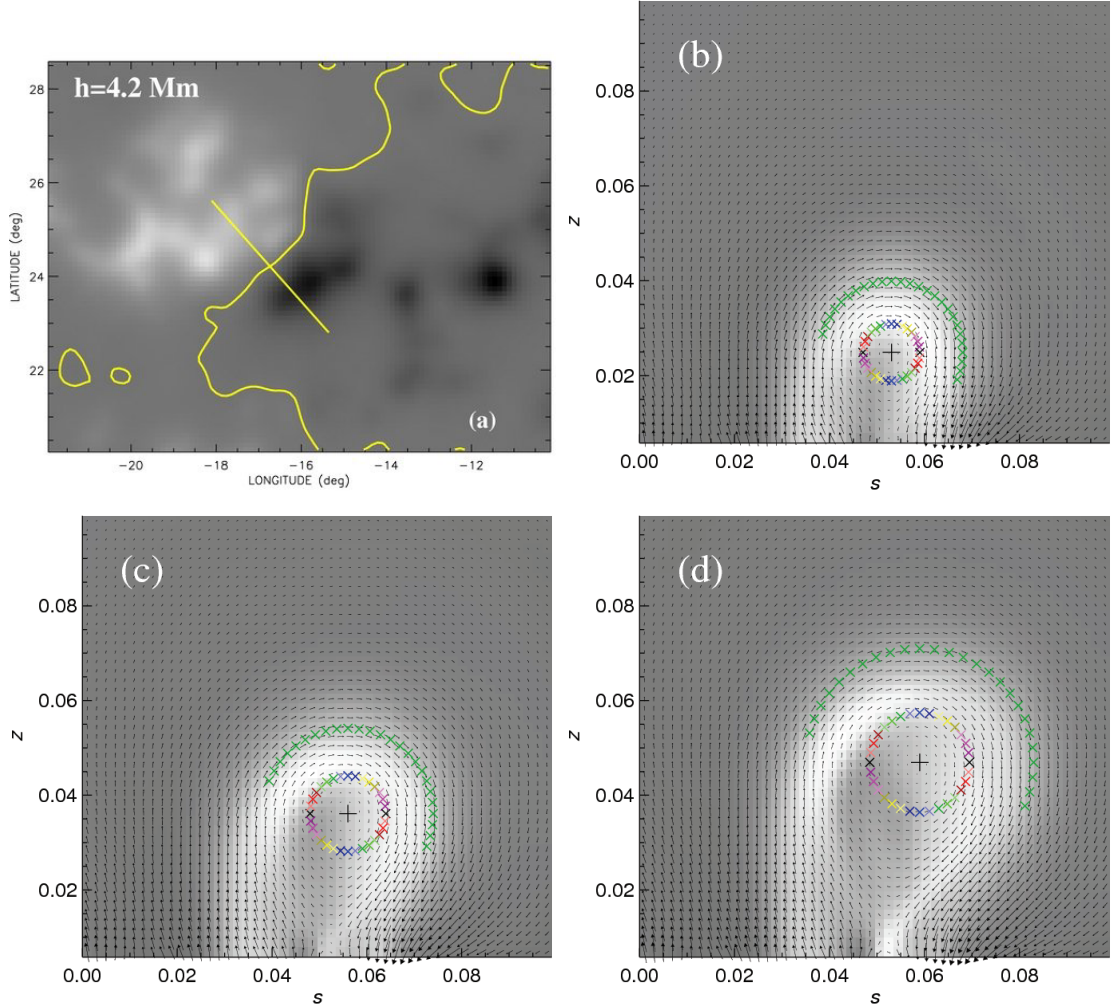}
\caption{\label{f:FRI}
 The three configurations investigated in this paper after 30\,000 MFR
 iteration steps. Shown are: 
 (a) the magnetogram $B_z(x,y)$ at a height of $z=4.2$~Mm,
     the PIL at this height,
     and the position of a vertical plane used in the analysis; the
     plane is placed perpendicular to the PIL such that it passes very
     near the apex point of the rope's magnetic axis for all
     configurations.
 (b)--(d) Force-free parameter $\alpha$ and in-plane magnetic field
     vectors in this diagnostics plane (for $z\ge4.2$~Mm; $s$ denotes
     horizontal distance). The configurations possess axial flux of 4,
     5, and $6\times10^{20}$~Mx for (b)--(d), respectively, and an
     identical poloidal flux of $10^{10}$~Mx~cm$^{-1}$. Lengths are
     given in $R_\odot/f$, i.e., they would correspond to a unit of
     $1R_\odot$ if scaled back to the original magnetogram size.
     The gray background in these panels corresponds to
     $\alpha=0$, and in the area shown $\alpha$ lies in the range
     $[-9,141]$, $[-8,125]$, and $[-8,118]$ for panels (b)--(d),
     respectively.
     Crosses show the initial position of the fluid elements used as
     start points of the field line tracing in the subsequent plots
     and animations; the trajectories of these fluid elements are
     followed through the simulations. Rainbow-colored crosses are
     placed inside the FR near its edge, and green crosses are placed
     in the inner part of the overlying flux.
     The plus signs mark the apex point of the flux rope's magnetic
     axis. The motion of this fluid element is used to generate the
     rise profiles in Figures~\ref{f:a4p1n_cfl}, \ref{f:a5p1n_cfl},
     and \ref{f:a6p1n_cfl}.
 (A color version of this figure is available in the online journal.)
}
\end{figure*}

The next step is to apply 30,000 iterations of MFR to each of the
three models. This drives the magnetic field toward a force-free state
(if one exists) or causes the field to slowly expand (if the
configuration is unstable). Each 3D magnetic model is then resampled
on the grid used by the MHD code. This grid is nonuniform in
longitude, latitude and height; it provides high spatial resolution in
the center of the AR and gradually reduced resolution
farther away from the center. Figure~\ref{f:FRI} shows vertical
cross-sections through the three models after resampling on the
Cartesian grid. For consistency with the designation in Paper~I we
will refer to the axial flux values of the inserted flux ropes
corresponding to the magnetogram size before the spatial scaling,
$4 \times 10^{20}$, $5 \times 10^{20}$ and $6 \times 10^{20}$~Mx. The
first panel indicates on a magnetic map at height 4.2~Mm where the
vertical cross-sections are taken. The quantity plotted in the three
other panels is $\alpha = \bm{j \cdot B} / B^2$, which is a measure of
the parallel electric current. Note that the currents are concentrated
near the edge of the flux rope, and that the height of the flux rope
increases with axial flux. Higher axial flux implies higher total
current running along the flux rope, which in turn implies a stronger
upward force on the current-carrying flux. This force can be
described as a
repelling force between the flux rope current and its subphotopheric
image if the flux rope runs nearly parallel to the solar surface
\citep{Kuperus&Raadu1974} and, equivalently, as the hoop force on
the rope if the rope arches upward considerably
\citep{Demoulin&Aulanier2010}. Consequently, the equilibrium height
increases with increasing axial flux. $\alpha$ is positive across
essentially the whole cross section of the flux rope, i.e., the field
is right handed and the rope carries nearly no return current.
Both models with $\Pa\ge5\times10^{20}$~Mx possess an
X-type magnetic structure---an HFT---running at low
height under the flux rope (see Paper~I for detail).

By applying the MFR much longer, \citeauthor{Su&al2011} concluded in
Paper~I that, for the given value of $\Fp=10^{10}$~Mx~cm$^{-1}$, the
model with $\Pa = 5 \times 10^{20}$~Mx lies near the stability
boundary in the $\Pa$--$\Fp$ plane. All configurations with
$\Pa \le 4.5 \times 10^{20}$~Mx relaxed to an apparently stable and
approximately force-free state in 30,000 MFR iterations, while all
configurations with $\Pa \ge 6 \times 10^{20}$~Mx did not relax,
rather the inserted flux rope continued to rise.


\section{Magnetohydrodynamic Modeling}
\label{s:MHD}

In the MHD modeling, the configurations obtained after 30,000
magnetofrictional iterations are used as the initial condition,
$\bm{B}_0(\bm{r})$, for the numerical integration of the compressible
ideal MHD equations in the zero-beta limit. We set the initial
velocity to  zero, $\bm{u}_0=0$, and choose
$\rho_0(\bm{r})=B_0(\bm{r})^{3/2}$ as model for the initial density.
This implies a gradual decrease of the initial Alfv\'en velocity
$V_\mathrm{A0}=B_0(2\mu_0\rho_0)^{-1/2}$ with distance from the strong
field in the core of the active region, such that the height profile
$V_\mathrm{A0}(z)$ approximately follows the empirical height profile
in \citet[][their Figure~5]{Vrsnak&al2002}. Such a choice of
$\rho_0(\bm{r})$ generally yielded good quantitative matches with
observed CME rise profiles in previous simulations
\cite[e.g.,][]{Torok&Kliem2005, Schrijver&al2008b, Kliem&al2012}.

The equations used and the numerical scheme are detailed in
\citet{Torok&Kliem2003}. The condition $\beta=0$ decouples the energy
equation from the system. Gravity can be neglected for both purposes
of the simulations in this paper: further relaxation of stable
force-free configurations and the study of CME acceleration in a
region of relatively strong magnetic field (an AR) in the
case of instability. Numerical diffusion breaks the frozen-in
condition when thin layers of inhomogeneous field (i.e., high current
densities) develop, thus allowing magnetic reconnection. The equation
of motion includes a viscous term solely to ensure numerical
stability.

\begin{figure*}[t]                                              
\centering
\includegraphics[width=.95\linewidth]{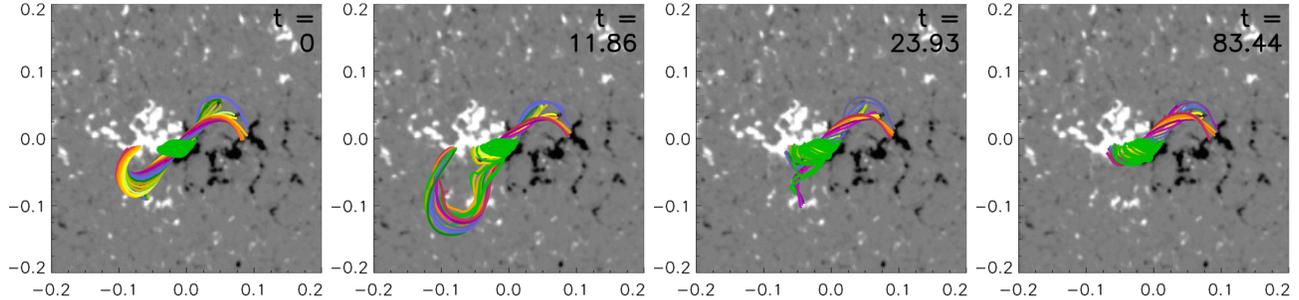}
\caption{\label{f:a4p1n_FR+overl}
 Stages in the relaxation of the model with $\Pa=4\times10^{20}$~Mx.
 Shown are field lines in vertical view and the magnetogram $B_z(x,y,0,t)$
 with the gray scale saturated at one tenth of the peak value.
 Rainbow-colored field lines lie in a flux shell slightly inside
 the boundary of the flux rope and green field lines visualize the
 inner part of the overlying flux (see Figure~\ref{f:FRI}).
 (A color version of this figure is available in the online journal.)
}
\end{figure*}

Careful attention is paid to the amount of numerical diffusion in the
integration. We use a modified verson of the two-step Lax-Wendroff
scheme that replaces the stabilizing but very diffusive Lax term in
the auxiliary step of the iteration by so-called artificial smoothing
\citep{Sato&Hayashi1979, Torok&Kliem2003}. While the Lax term replaces
the value of the integration variable by the average of the values at
the six neighboring grid points, the artificial smoothing replaces
only a fraction of the value, $\sigma\ll1$, by the average at the
neighboring points. For the magnetic field, this averaging is switched
off completely in the relaxation runs ($\sigma_B=0$) to minimize any
slow diffusive change of the force-free equilibrium, improving the
convergence toward the equilibrium. The eruption of the unstable
configuration can be followed with the same setting; however, a
parametric study of the field smoothing yields the most reliable
numerical results and best numerical stability if a small level of
smoothing, $\sigma_B\sim10^{-3}$, is adopted in the volume under the
rising flux rope, where a vertical current sheet develops (see
Section~\ref{s:eruption} for detail). This operation is not applied in
the bottom layer of the box, $\{z=0\}$, so that there is no diffusion
of the field in the magnetogram plane.

Minimizing the numerical diffusion in the momentum equation allows us
to capture any residual forces that result from our initial conditions
being out of equilibrium. We have experimented with the values of
$\sigma_u$ and $\nu$ (the coefficient of viscosity), and chosen them
as close to the limit of numerical stability as reasonably possible,
i.e., a lowering of either of them by a factor of 2 would lead to
numerical instability in the course of the simulation run. Best
numerical stability is obtained by smoothing the density at the same
level. In the main volume of the box $\sigma_\rho=\sigma_u=0.005$. A
gradual enhancement of the smoothing toward the bottom boundary, by a
factor two, is necessary to allow stable integration in the presence
of the strong flux gradients in the magnetogram and their associated
currents. Uniform small viscosity is chosen: $\nu=0.002$ (after
normalization).

The integration volume corresponds to
$0.72R_\odot\times0.72R_\odot\times1R_\odot$ when the magnetogram is
scaled back by the factor $f$ to the original edge length. This volume
is discretized by a Cartesian grid with uniform resolution of
$0.002R_\odot$ (1.9~arcsec) in the AR and its immediate
surroundings ($|x|<0.1R_\odot$, $|y|<0.1R_\odot$, $z<0.2R_\odot$) and
a gradually increasing spacing in the outer parts, reaching
$0.0065R_\odot$ at the side boundaries and $0.015R_\odot$ at the top.
Closed boundaries are implemented at the sides and top of the volume
by keeping the velocity at zero, including the first inner grid layer,
so that the field vector in the boundary layers does not change. At
the bottom boundary the velocity is kept at zero in the magnetogram
plane, $\bm{u}(x,y,0,t)=0$, which ensures that the normal magnetogram
component $B_z$ keeps its initial values, but allows the transverse
components to change, either to approach an NLFFF, or to consistently
evolve with an eruption.

The box height ($1R_\odot$) is chosen as the length unit. The field
strength is normalized by the peak value, $\max(B_0)=955$~Gauss, which
lies in the magnetogram plane. Velocities are normalized by the peak
value of the initial Alfv\'en velocity, $\hat{V}_\mathrm{A0}$, whose
location coincides with the location of the highest field strength.
The initial Alfv\'en velocity in the body of the flux rope is about
half this value. Time is normalized by the corresponding Alfv\'en time
$\tau_\mathrm{A}=R_\odot/\hat{V}_\mathrm{A0}$.

\section{Relaxation of Stable and Nearly Marginally Stable Configurations}
\label{s:relaxation}

\begin{figure*}[t]                                               
\centering
\includegraphics[width=.95\linewidth]{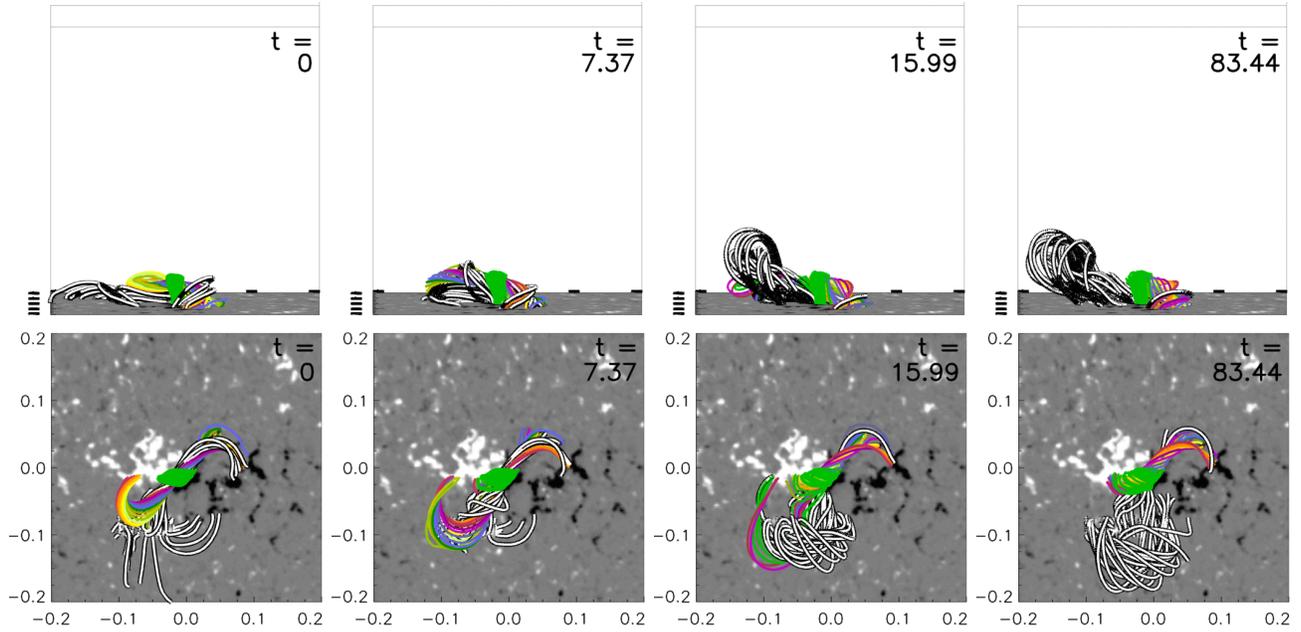}
\caption{\label{f:a4p1n_FR+overl+pospatch}
 Same as Figure~\ref{f:a4p1n_FR+overl} but with additional field lines
 traced from a set of fixed points located at $z=0$ in the minor
 positive flux concentration in the southeast of the active region.
 The strong interaction of this flux with the inserted rope leads to
 the southward bulging of the rope and to multiple reconnections
 involving the splitting and the subsequent recovery of a single rope.
 The upper row shows the cube above the selected magnetogram area in
 perspective views oriented in $-x$ direction.
 (A color version and an animation of this figure are available in the
 online journal.)
}
\end{figure*}

\begin{figure}[t]                                                
\centering
\includegraphics[width=\linewidth]{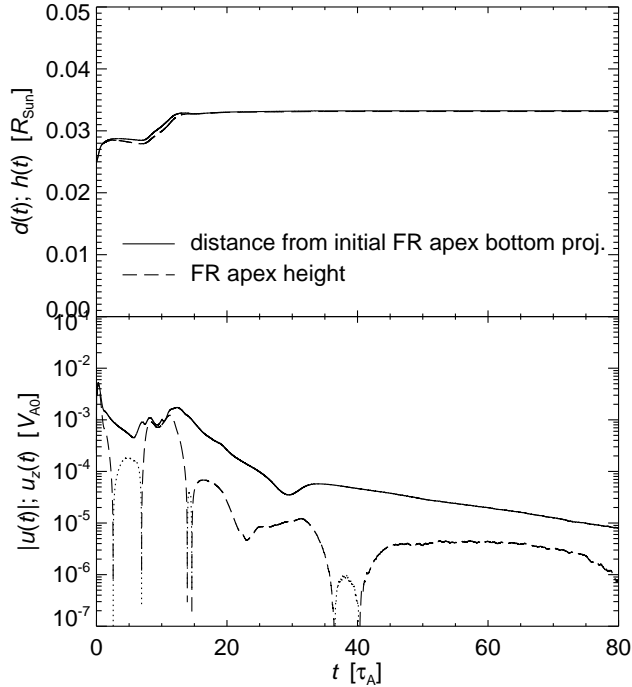}
\caption{\label{f:a4p1n_cfl}
 Height-time and velocity-time profile showing the relaxation of the
 model with $\Pa=4\times10^{20}$~Mx. The fluid element initially
 located at the estimated apex point of the flux rope's magnetic axis,
 shown in Figure~\ref{f:FRI} by a plus sign, is followed. The solid
 lines show the distance $d(t)$ from the initial projection of the
 apex onto the bottom plane and the 3D velocity of the fluid element,
 $|\bm{u}(t)|$. The dashed lines show the height above the bottom
 plane and $u_z(t)$. Dotted lines in the logarithmic plot show
 downward velocities ($|u_z(t)|$ for $u_z<0$).
}
\end{figure}

In our MHD description both models with $\Pa\le5\times10^{20}$~Mx are
found to relax to a stable force-free equilibrium (NLFFF). As
expected, the relaxation proceeds faster and deeper for the model with
$\Pa=4\times10^{20}$~Mx; nevertheless, the initial configuration
clearly includes a small level of residual forces and the flux rope
experiences further reconnection with the ambient field in the course
of the relaxation. The overall evolution is characterized in
Figure~\ref{f:a4p1n_FR+overl}. From the field line plot at $t=0$ it is
clear that the eastern part of the inserted flux rope has evolved
significantly in the course of the MFR: the flux connected from the
end point of the path shown in Figure~\ref{fig2} to the main positive
polarity in the northeast of the region (i.e., in the direction of
loop set~2 in Figure~\ref{fig1}) now bulges out to the south in the
direction of loop set~3. Close similarity to loop set~1 on the western
side was achieved with relatively little change. The strong evolution
of the eastern part continues in the course of the MHD relaxation,
driven primarily by reconnection with the minor positive polarity in
the southeast of the region. This wraps flux from the minor polarity
around the flux rope, passing northward under the rope and from there
arching over the rope toward the main negative polarity in the west.
The bulging of the southern part of the rope is thus considerably
enhanced. As a consequence, the flux rope reconnects with the
perturbing flux and its whole eastern elbow splits.

This interaction with the ambient flux rooted in the minor positive
polarity is shown in more detail in Figure~\ref{f:a4p1n_FR+overl+pospatch}
($t=0\mbox{--}16\tau_\mathrm{A}$)
and its accompanying animation. Further reconnection in this
area subsequently simplifies the structure of the inserted flux which
returns to an essentially unsplit flux rope, shorter on the eastern
side, where it now displays closer similarity in shape with loop
set~2 (Figures~\ref{f:a4p1n_FR+overl}--\ref{f:a4p1n_FR+overl+pospatch}
at $t>20\tau_\mathrm{A}$).
New low-lying connections run from the minor positive polarity
to the eastern part of the main negative polarity under the main body
of the relaxed flux rope (Figure~\ref{f:a4p1n_FR+overl+pospatch}
at $t=83\tau_\mathrm{A}$); these are similar to loop set~3. Note that
these connections differ from the ones at $t=0$. Their existence is
not even indicated in the potential field. Additionally, new
high-arching field lines extend from the minor positive polarity with
a dominant east-west orientation. They have absorbed some of the twist
of the inserted flux. One can speculate that such higher field lines
confine plasma of lower density, so that these connections should be
less visible in X-ray and EUV images. Some irregularly shaped emission
does exist in this area at the temperature ($T\approx1.5$~MK) sampled
by the AIA 193~{\AA} channel (Figure~\ref{fig1}(b)). It consists of
moss essentially cospatial with the minor flux concentration and of
weak diffuse emission slightly southward, which has a dominantly
east-west orientation reminiscent of the high-arching field lines seen
in the simulation from $t\sim15\tau_\mathrm{A}$ onward.

\begin{figure*}[t]                                               
\centering
\includegraphics[width=.95\linewidth]{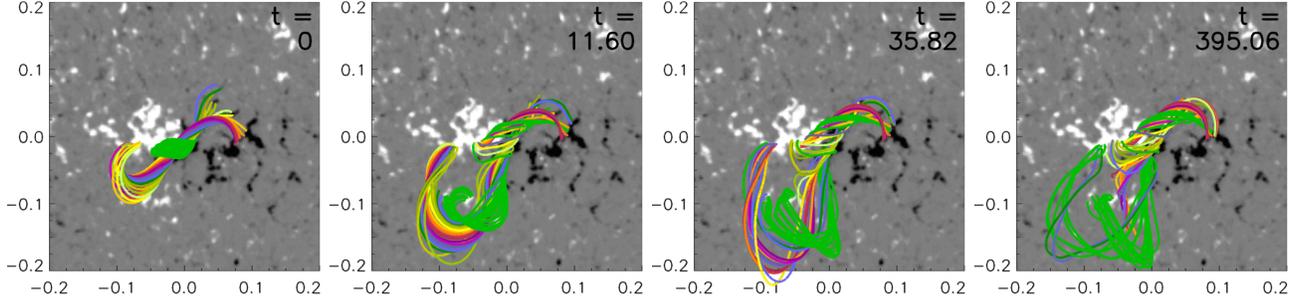}
\caption{\label{f:a5p1n_FR+overl}
 Stages in the relaxation of the model with $\Pa=5\times10^{20}$~Mx in
 the same format as Figure~\ref{f:a4p1n_FR+overl}.
 (A color version and an animation of this figure are available in the
 online journal.)
}
\end{figure*}

More quantitative information about the relaxation, displayed in
Figure~\ref{f:a4p1n_cfl}, is obtained by monitoring the position and
velocity of the fluid element initially at the apex point of the
rope's magnetic axis (marked by a plus sign in Figure~\ref{f:FRI}). We
first note a minor but quick shift in apex height at the very
beginning of the run ($t\lesssim2\tau_\mathrm{A}$), which potentially
results from two effects. One reason lies in the difference between
the codes in obtaining the current density. While the MFR code runs on
a staggered grid, the MHD code iterates all variables at the same
positions, leading to differences in derived quantities between the
codes which increase with degrading spatial resolution. Since all
models include a relatively thin layer of enhanced current density at
the edge of the flux rope which is resolved by only $\sim5\mbox{--}10$
grid cells (Figure~\ref{f:FRI}), the Lorentz force differs somewhat
between the MFR and MHD descriptions in this critical layer, and so
does the equlibrium height of the flux rope. It is clear, however,
that the overall difference in the volume of the flux rope must be
moderate at the resolution employed: the height of the flux rope's
apex increases by only 15\% (from $h_0=0.025$ to $h(t\!=\!2)=0.029$).
Stronger differences exist in the first few grid layers above the
magnetogram, where the strong fragmentation of the flux is not fully
resolved, resulting in small volumes of high current density. Although
the associated Lorentz forces do not have a coherent direction, they
may still contribute to the change in equilibrium height. This
contribution can be estimated through a comparison with the relaxation
behavior of the potential field, calculated on the spherical grid of
CMS and resampled on the Cartesian grid of the MHD code in the same
manner as the three AR models. The fluid element at the
same initial position as the one considered in
Figure~\ref{f:a4p1n_cfl} rises by only 2.1\% to $h=0.0255$, also
experiencing most of this displacement (80\%) within the first two
Alfv\'en times. Thus, the contribution from discretization errors in
the current density near the bottom of the box appears to be only
minor. Second, part of the initial rise may be due to the status of
relaxation after the 30,000 MFR iterations. Although already deep, as
the subsequent rapid decrease of the velocities shows, it is obviously
not yet complete (see above). The low level of viscosity and velocity
smoothing in the MHD relaxation allows the small residual forces to
build up a noticeable velocity in the motion to the stable equilibrium
height.

Subsequently, a further rise to $h=0.033$ occurs which is associated
with the reconnections discussed above. The velocity of the fluid
element shows a nearly monotonic decrease, with a weak transient
enhancement during $t\sim(5\mbox{--}20)\tau_\mathrm{A}$ related to the
southward bulging of the flux rope. Following the initial jump (i.e.,
after $t\approx2\tau_\mathrm{A}$), the vertical velocity falls by more
than three orders of magnitude to $u_z\sim10^{-6}~\hat{V}_\mathrm{A0}$,
which is a clear signature that the flux rope relaxes deeply.
To further quantify the relaxation, we compute the
current-weighted average sine of the angle between current and
magnetic field, $\sigma_j=\sum_ij_i\sigma_i/\sum_ij_i$, where
$\sigma_i=|\bm{j}_i\bm{\times B}_i|/(B_ij_i)$ and the index $i$ runs
over all grid cells. Following standard practice in the use of this
quantity in magnetogram extrapolations \cite[e.g.,][]{DeRosa&al2009},
the outer parts of the box are excluded. Additionally, we exclude the
first three grid layers at the bottom of the box, where the strong
fragmentation of the flux is not fully resolved. Their contribution
raises the value of $\sigma_j$ by about an order of magnitude, while
the outer parts of the box contribute only a further factor of
$\approx\!1.2$. In the volume $|x|<0.2$, $|y|<0.2$, $0.006<z<0.4$ we
find $\sigma_j(0)=0.066$ and $\sigma_j(t\!=\!83)=0.015$,
falling monotonically, except for a moderate enhancement during
$t\sim(5\mbox{--}20)\tau_\mathrm{A}$. Here it should be kept in mind
that the initial degree of force freeness, $\sigma_j(0)$, is degraded
from the MFR relaxation result by the transition from the staggered
CMS grid to the non-staggered grid of the MHD code. The
magnetic energy in the box decreases by 2.7\% in the course of the
relaxation.

\begin{figure}[t]                                                
\centering
\includegraphics[width=\linewidth]{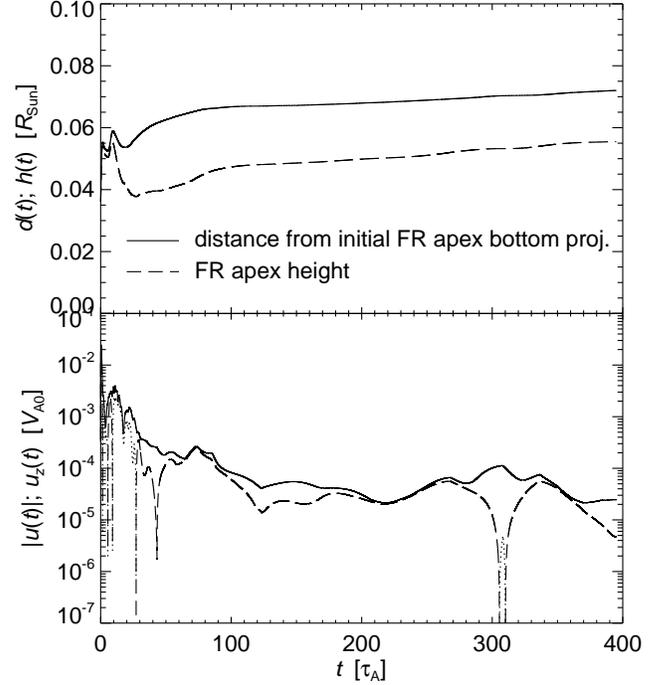}
\caption{\label{f:a5p1n_cfl}
 Height-time and velocity-time profile showing the relaxation of the
 model with $\Pa=5\times10^{20}$~Mx in the same format as 
 Figure~\ref{f:a4p1n_cfl}.
}
\end{figure}

The model with $\Pa=5\times10^{20}$~Mx follows a rather similar
evolution in the course of its relaxation, albeit on a much longer
time scale and relaxing somewhat less deeply
(Figures~\ref{f:a5p1n_FR+overl} and \ref{f:a5p1n_cfl}). In particular,
the reconnection of the inserted flux rope with the minor positive
flux concentration in the southeast of the AR yields a
similar bulging, splitting, and subsequent trend of recovery. The
resulting relaxed state is largely similar, but here the flux rope
remains partly split. Again, the major part of the relaxed rope
matches the observed loop sets~1 and 2 relatively well. The split-off
part of the rope and some of the new connections between the minor
positive polarity and the main negative polarity correspond well to
loop set~3. Finally, diffuse, high-arching loops, with a dominant
east-west orientation, are formed south of the minor positive
polarity, as in the model with less axial flux. A noteworthy
difference is the higher initial displacement of the flux rope by
$\approx50\%$ from $h_0=0.036$ to $h(2)=0.053$ in the first
$\approx2\tau_\mathrm{A}$. It can be seen in Figure~\ref{f:FRI} that
the resolution of the current layer at the surface of the flux rope in
the two models is rather similar, so the stronger initial displacement
must be primarily due to a less complete relaxation in the course of
the MFR. Indeed, we find $\sigma_j(0)=0.074$, slightly higher than
for the model with $\Pa=4\times10^{20}$~Mx. Except for a moderate
enhancement during $t\sim(5\mbox{--}15)\tau_\mathrm{A}$, the average
angle between $\bm{j}$ and $\bm{B}$ decreases monotonically to
$\sigma_j(t\!=\!83)=0.019$ and $\sigma_j(t\!=\!395)=0.012$.

After about $100\tau_\mathrm{A}$ the residual velocities lead to a
very slow rise of the flux rope position, hardly visible in the field
line plots but obvious in Figure~\ref{f:a5p1n_cfl}. It is difficult to
judge whether this evolution is part of the relaxation or due to of a
diffusive drift of the equilibrium in the long computation comprising
of slightly over $10^6$ iteration cycles. Animated field line plots
(Figure~\ref{f:a5p1n_FR+overl}) show that the rope continues to evolve
over this long time period (some of the field lines change their
photospheric connections considerably), and on average the velocity of
the rope continues to decrease (apart from episodic enhancements);
both properties suggest that the slow drift may be part of the
relaxation. Nevertheless, it is unclear whether continuing this run
can lead to further understanding of this model free from numerical
effects, so the run was terminated. Adopting the scaling of the
simulation in Section~\ref{ss:scaling}, the nearly 400 Alfv\'en times
of this relaxation correspond to up to 2.3~days, longer
than the equilibrium can be assumed to be static. By the end of the
relaxation, the magnetic energy in the box has dropped by 4.5\%. The
model appears stable in the MHD treatment, but is doubtlessly much
closer to the point of marginal stability than the previous model.

\begin{figure}[t]                                                
\centering
\includegraphics[width=\linewidth]{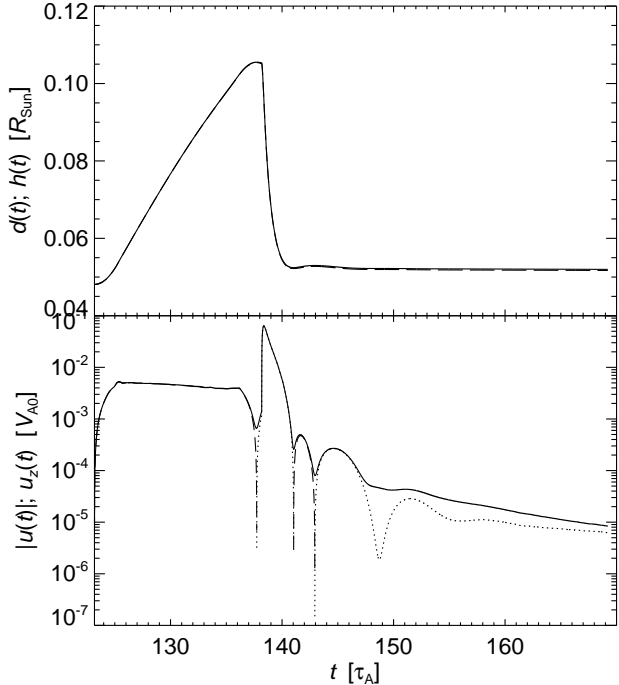}
\caption{\label{f:a5p1n_pert_cfl}
 Evolution of the model with $\Pa=5\times10^{20}$~Mx following a
 strong perturbation applied at a stage of deep relaxation
 ($t=123\tau_\mathrm{A}$). Position and velocity of the fluid element
 at the apex of the flux rope are plotted in the same format as in
 Figure~\ref{f:a5p1n_cfl}, with the 3D position $d(t)$ here
 referenced to the bottom projection of the apex point at
 $t=123\tau_\mathrm{A}$.
}
\end{figure}

To further test for the stability of this model, we have perturbed it
at the times of apparently deepest relaxation, $t=123\tau_\mathrm{A}$
and $t=395\tau_\mathrm{A}$. The flux rope was raised to
$h\approx0.105$ by prescribing upward velocities at the apex in a
sphere of radius equal to the minor flux rope radius for
$15\tau_\mathrm{A}$. Slightly below this height the unstable third
model starts its fast ascent (see next section). Despite the very
strong distortion, the flux rope returned close to the original
position, executing quickly decaying relaxation oscillations. This is
shown in Figure~\ref{f:a5p1n_pert_cfl} for the perturbation applied at
$t=123\tau_\mathrm{A}$. Nearly identical behavior results for the
perturbation applied at the end of the relaxation run. These tests
support the interpretation that the model with axial flux
$\Pa=5\times10^{20}$~Mx relaxes to a stable equilibrium.

The model can be driven to eruption by further flux cancellation. This
was tested by prescribing horizontal flows converging toward the PIL
in the strong-flux section of the PIL and allowing for diffusion of
the field in the vicinity of the PIL, analogous to the diffusion in
\citet{Aulanier&al2010} and \citet{Amari&al2011}. Cancellation of about
10\% of the unsigned flux in the region caused the eruption of the
flux rope (but more extended experimenting might find eruption at an
earlier stage). Different from observation, the flux rope took a
nearly vertical path. This and the fact that an appropriate modeling
of pre-eruption driving by flux cancellation should start from a
magnetogram taken at an earlier time, led us to reserve a more
detailed investigation of such simulations for a future study.


\section{Eruption of the Unstable Configuration}
\label{s:eruption}

\begin{figure*}[t]                                              
\centering
\includegraphics[width=.78\linewidth]{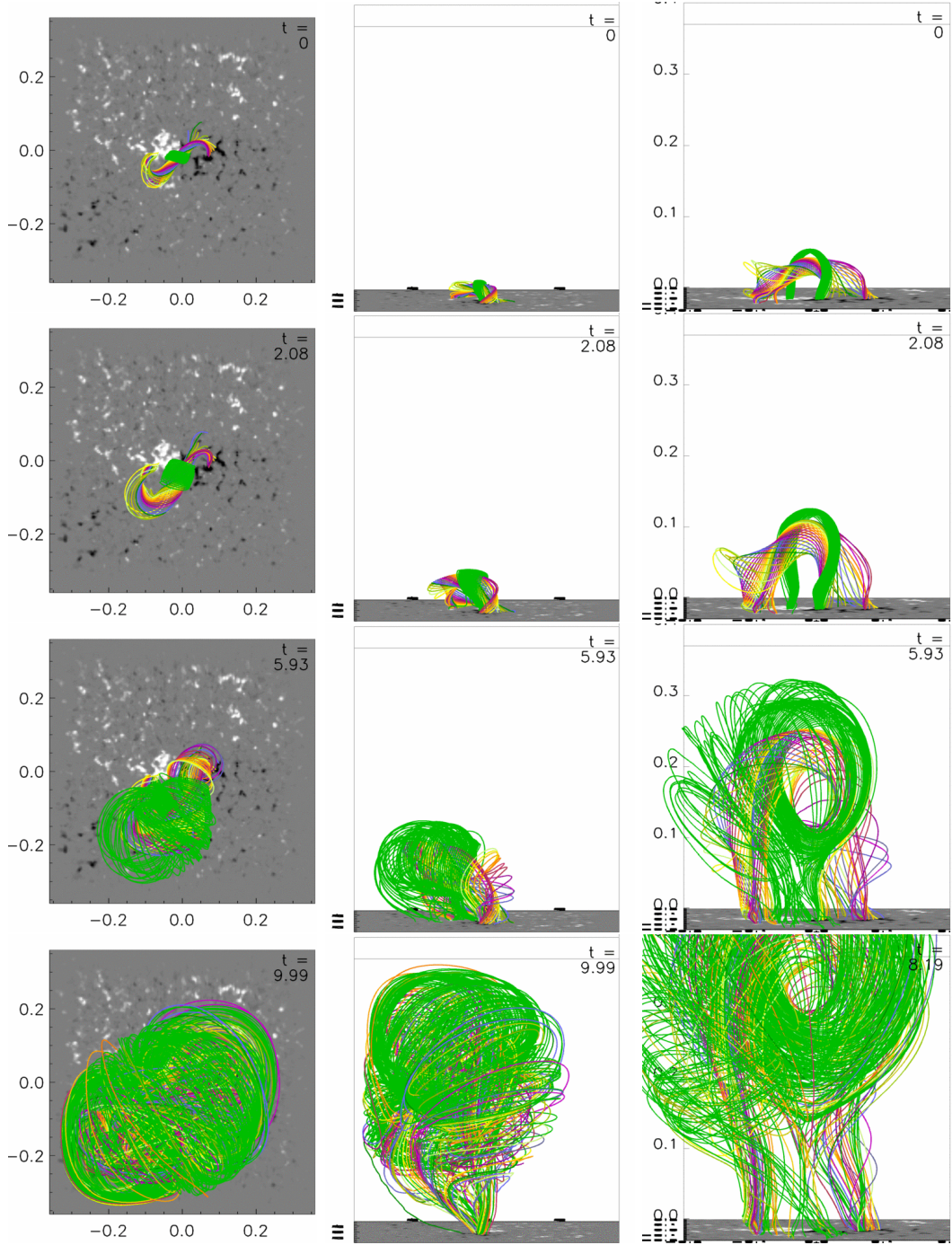}
\caption{\label{f:a6p1n_FR+overl}
 Flux rope eruption in the model with $\Pa=6\times10^{20}$~Mx.
 The format is similar to Figures~\ref{f:a4p1n_FR+overl} and
 \ref{f:a5p1n_FR+overl}. The whole box is shown in the left two
 columns. The eruption starts strongly inclined, as observed.
 Subsequently, the erupting flux is deflected at the closed side
 boundary. Originally overlying flux (green field lines) is strongly
 reconnected into the rope by flare reconnection in the vertical
 current sheet that forms under the rope (see
 Figure~\ref{f:a6p1n_vcut}).
 (A color version and an animation of this figure are available in the
 online journal.)
}
\end{figure*}

Similar to the result of the MFR in Paper~I, the rope with axial flux
$\Pa=6\times10^{20}$~Mx does not relax to an equilibrium, but rather erupts.
As expected, the initial configuration deviates stronger from
force freeness than the two stable models, $\sigma_j(0)=0.083$.
The evolution starts with a rapid displacement of the rope within the
first two Alfv\'en times, comprising the lifting of the middle part
and the southeastward expansion of the eastern elbow, which are very
similar to the initial behavior in the relaxation runs of the previous
section. The interaction with the flux rooted in the minor positive
polarity is also similar but does not lead to a split rope (which
occurs at $t>5\tau_\mathrm{A}$ in the relaxation runs) because of the
commencing eruption.

A seamless transition to an initially slower but accelerating rise
follows, leading to the eruption of the whole flux rope (see
Figure~\ref{f:a6p1n_FR+overl} and the accompanying animation). The
main direction of this motion is upward and southward, initially (at
$t=2\tau_\mathrm{A}$) at an inclination of 40$^\circ$--45$^\circ$ to
the vertical, which is very close to the observed direction.
Interaction with a nearby coronal hole \citep{Gopalswamy&al2009} and
asymmetry of the photospheric flux distribution
\citep{Panasenco&al2013} have been proposed to cause deviations from
radial ascent. The former is not included in our model, while the
latter effect is present, but presumably not very strong: the main
positive polarity in the AR contains only $\approx10\%$
more flux than the main negative polarity. However, the positive
polarity is more compact. An additional effect is suggested by the
shape of loop sets~1 and 3: a southward-directed slingshot action by
their common flux after the flux rope loses equilibrium. The fact that
the dimmings of the eruption begin to develop near their end points
(Figure~\ref{fig1}(d)) indicates that these loop sets possess common
flux which plays a role in the eruption.
The shape of this flux implies a southward tension force in
addition to the upward Lorentz force which drives the eruption.
Different from observation,
the rise soon begins to gradually turn more radial (at
$t=6\tau_\mathrm{A}$ the inclination is $\approx30^\circ$). This may
be due to the missing action of the polar coronal hole and due to the
slingshot effect being weaker than in reality, since the time for loop
set~3 to develop fully is not available in the evolution of the
unstable model. The latter aspect may be improved by applying longer
MFR in the preparation of this model.

The outer layers of the expanding flux rope begin to experience
compression at the side boundary of the computation box already at
$\sim6\tau_\mathrm{A}$, while the core hits the boundary with a delay
of $(1\mbox{--}2)\tau_\mathrm{A}$. The evolution up to this point
shows a number of similarities to the observed behavior (see also
below). Subsequently, the upward deflection of the rope is of course
unrealistic, but the continuing strong expansion of the minor flux
rope radius and the continuing reconnection under the rope, discussed
below, are likely qualitatively consistent with the modeled solar
event---until the top boundary is encountered.

\begin{figure*}[t]                                             
\centering
\includegraphics[width=.82\linewidth]{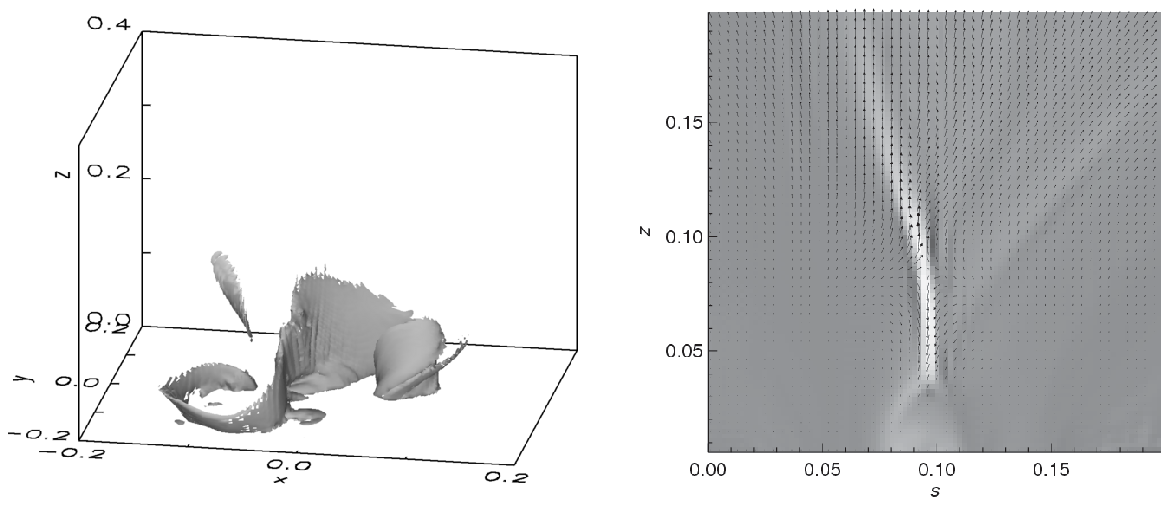}
\caption{\label{f:a6p1n_vcut}
 Vertical current sheet in the unstable configuration at
 $t=6.6\tau_\mathrm{A}$.
 (\emph{Left}): Isosurface of current density at $0.1\max(|\bm{j}|)$.
 (\emph{Right}): Force-free parameter $\alpha(s,z)$ and in-plane
 velocity vectors in the vertical cut plane shown in Figure~\ref{f:FRI}.
 $\alpha=0$ is plotted in gray, and the peak value (white) is
 $\max(\alpha)=483$ at $(s,z)=(0.092,0.096)$.
}
\end{figure*}

\begin{figure*}[t]                                              
\centering
\includegraphics[width=.82\linewidth]{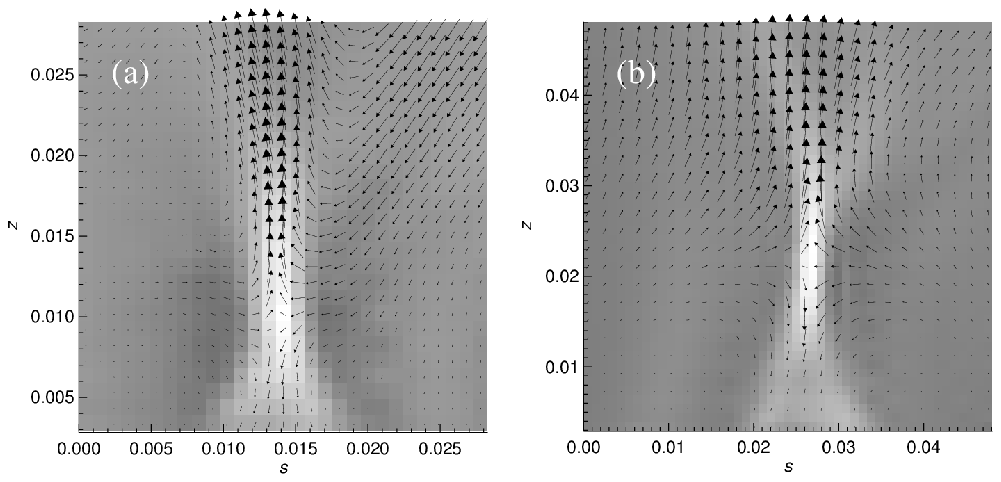}
\caption{\label{f:a5p1n+a6p1n_reconn}
 Reconnection flows of the models with
 (a) $\Pa=5\times10^{20}$~Mx and
 (b) $\Pa=6\times10^{20}$~Mx
 in the vertical cut plane shown in Figure~\ref{f:FRI}
 at $t=2\tau_\mathrm{A}$.
 The peak in-plane velocities are 0.0045 in (a) and 0.034 in (b).
 Additionally, $\alpha(s,z)$ is displayed as in
 Figure~\ref{f:a6p1n_vcut}, with
 $\max(\alpha)=330$ at $(s,z)=(0.014,0.010)$ in (a) and
 $\max(\alpha)=404$ at $(s,z)=(0.027,0.023)$ in (b).
}
\end{figure*}

The eruption of the rope lifts the overlying flux. This flux is
initially sheared (Figure~\ref{f:a6p1n_FR+overl}) but becomes more
antiparallel in the volume under the rope as it is lifted. The
corresponding induced current points primarily horizontally and is
associated with a Lorentz force pointing toward the essentially
vertical layer under the flux rope where the vertical field component
of the ambient flux changes sign. The induced current thus pinches
into the vertical current sheet, or flare current sheet, known from
the standard flare model. One can understand this pinching also as the
3D generalization of the well-known instability of a magnetic X point
in two dimensions. The generalized X-type structure is known as an
HFT. Its pinching into a current sheet
following an appropriate perturbation has been demonstrated by
\citet{Titov&al2003} and \citet{Galsgaard&al2003}. The initial
configuration contains such a structure (see Figure~\ref{f:FRI} and
Paper~I). Note that the pinching is solely driven by the Lorentz force
in our zero-beta simulation (the pressure gradient would additionally
contribute if $\beta>0$).

\begin{figure*}[t]                                              
\centering
\includegraphics[width=.67\linewidth]{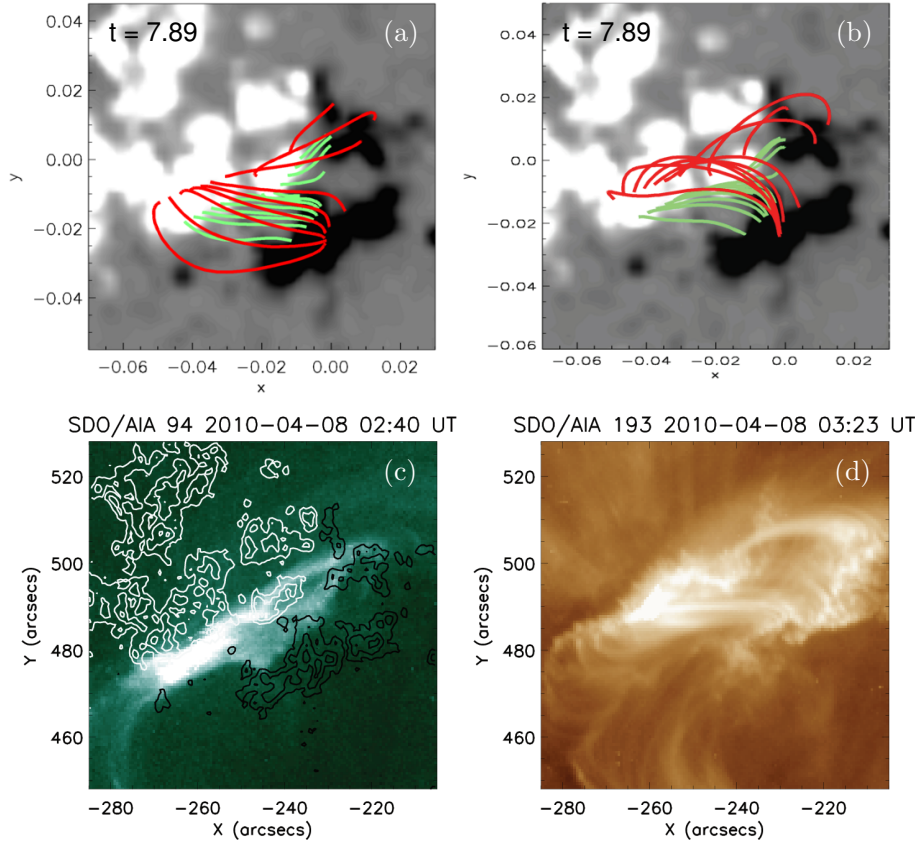}
\caption{\label{f:a6p1n_flareloops}
 Reconnected field lines in the unstable configuration in (a) vertical
 view and (b) perspective view inclined by $25^\circ$,
 overplotted on the magnetogram $B_z(x,y,0,t)$.
 (c--d) Flare loops of the event at the times corresponding to,
 respectively, $t=2.8\tau_\mathrm{A}$ and $t=7.9\tau_\mathrm{A}$ in
 the simulation. Contours of the LOS field component are included
 in (c).
 (A color version of this figure is available in the online journal.)
}
\end{figure*}

\begin{figure*}[t]                                             
\centering
\includegraphics[width=.82\linewidth]{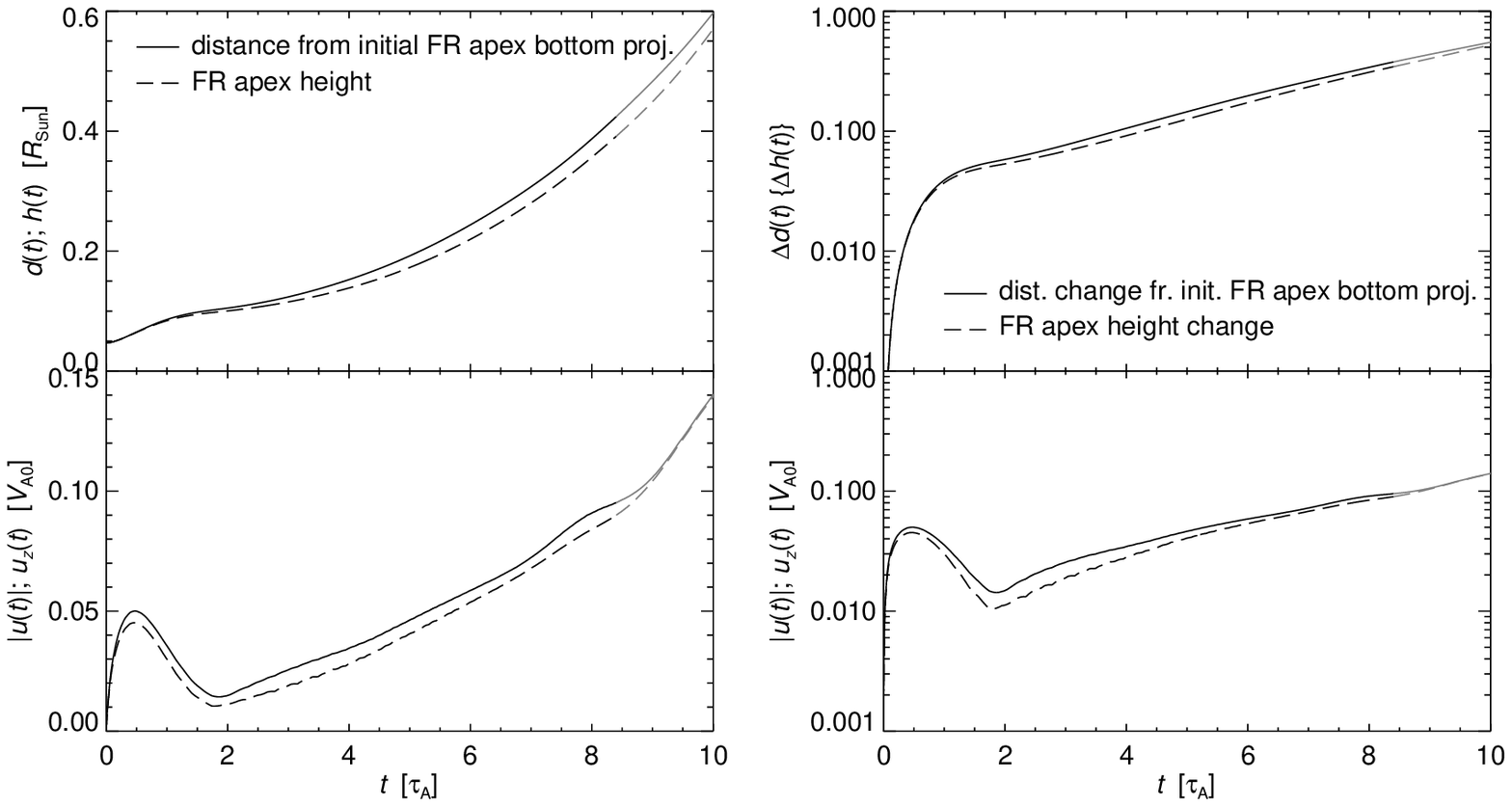}
\caption{\label{f:a6p1n_cfl}
 Rise profile of the flux rope apex in the unstable
 configuration, shown in the left panels in a format similar to
 Figure~\ref{f:a4p1n_cfl}. The right panels show the same information on
 a logarithmic scale. The section of the rise after reflection at the
 side boundary is plotted in gray.
}
\end{figure*}

Figure~\ref{f:a6p1n_vcut} displays the vertical current sheet that
forms under the rising flux rope. Since it is a true current sheet
\cite[with exponentially rising current density and correspondingly
decreasing width in ideal MHD;][]{Titov&al2003}, the
pinching process reaches saturation in a standard one-fluid MHD
description only if sufficient diffusion is provided. Otherwise, it
proceeds toward the limit of the employed numerical scheme, resulting
in unphysical filamentation of the current sheet \cite[which develops
on the grid scale and is not related to the plasmoid instability of
current sheets;][]{Leboeuf&al1982, Loureiro&al2007}. To verify the
eruption of the third model under uniform numerical settings, and as a
reference, we have first followed the latter evolution in a run
without magnetic field smoothing, $\sigma_B=0$. Magnetic reconnection
develops in the vertical current sheet and forms coherent reconnection
outflows in spite of the ensuing filamentation, so the main
ingredients of an eruption---loss of equilibrium and magnetic
reconnection---are present in spite of the numerical artefacts.
Repeating this run with magnetic field smoothing, chosen to be uniform
in the first instance, we find that the flux rope rise velocity is not
strongly influenced as long as $\sigma_B\lesssim10^{-3.5}$; it then
stays within 20\% of its value for $\sigma_B=0$. Beyond that
level, the rise velocity falls off quickly, and for
$\sigma_B\ge10^{-2.5}$ all velocities diffuse away after the initial
upward displacement of the rope. It is clear that when the magnetic
diffusion $\sigma_B$ is large, the currents and Lorentz forces in the
body of the flux rope are weaker, artificially stabilizing the
solution. The filamentation of the current sheet decreases with
increasing $\sigma_B$ but disappears only for
$\sigma_B\gtrsim10^{-3}$. Therefore, the appropriate numerical setting
for the present model consists in applying the smoothing of the
magnetic field only in the volume under the flux rope using a level of
$\sigma_B=10^{-3}$, with a gradual transition to $\sigma_B=0$ at the
bottom and at the sides. This avoids the filamentation of the vertical
current sheet and yields a rise velocity of the flux rope apex very
close to (slightly higher than) the value obtained in the absence of
the smoothing. Figures~\ref{f:a6p1n_FR+overl}--\ref{f:criterion} show
the results of this run.

%

In addition to the pinching of the HFT under the rising flux rope,
part of the current layer at the periphery of the rope steepens, while
the main part of the current layer weakens (since the total current
through the rope must decrease to power the eruption). The bottom part
of the current layer at the side of the flux rope facing the stronger
ambient field northeastward of the PIL builds up current densities
comparable to the ones in the pinched HFT underneath; thus, the
vertical current sheet extends upwards asymmetrically along this side
of the flux rope.

The flows triggered by the eruption are shown in the second panel of
Figure~\ref{f:a6p1n_vcut}. One can see the rise and expansion velocity
in the whole cross section of the flux rope, as well as in the ambient
volume, which is threaded by field lines lifted by the rope. The
southward inclination of the eruption is apparent. Reconnection flows
develop in the vertical current sheet essentially simultaneous with
the rise of the rope. The first indication of the downward reconnection
outflow becomes visible at $t\approx1\tau_\mathrm{A}$, i.e., already
in the initial upward displacement to the rope's equilibrium height.
As the eruption progresses, the reconnection flows evolve joint with
the vertical current sheet; thus, they become aligned with the
northeastward periphery of the rope in the main phase of the eruption
shown in the figure. This asymmetry may contribute to the inclination
of the rise direction \cite[consistent with the findings
of][]{Panasenco&al2011}. It may also yield a contribution to the roll
effect \citep{Panasenco&al2011, Su&vanBallegooijen2013},
which is weakly indicated by animated
field line plots from this simulation, as well as in the EUVI-A data
shown below in Figure~\ref{f:a6p1n_scaling}. Finally, we note that the
upward reconnection outflow reaches slightly higher velocities than
the flux rope and its immediate surroundings, so that the new flux
carried into the rope is likely to have a direct positive effect on
the acceleration in the main body of the rope.

It is worth noting that the stable model with $\Pa=5\times10^{20}$~Mx
also possesses an HFT (see Figure~\ref{f:FRI}(c) above and Figure~7(d)
in Paper~I). This HFT pinches as well into a short vertical current
sheet during the initial upward displacement, and reconnection
commences as early as for the unstable model; see the direct
comparison of the models at $t=2\tau_\mathrm{A}$ (i.e., after the
initial displacement) in Figure~\ref{f:a5p1n+a6p1n_reconn}. However,
an eruption does nevertheless not occur. This indicates that
reconnection in the vertical current sheet of the considered coronal
NLFFF models is not able to drive an eruption by itself, but rather
that the driving by an unstable flux rope is required. A downward
return flow is induced in the stable case by the rapid initial rise of
the flux rope (Figure~\ref{f:a5p1n+a6p1n_reconn}(a)); this would not
occur in a quasi-static evolution. This flow is seen to join the
reconnection inflow, so it does not appear to work against the further
development of reconnection in this model.

Figure~\ref{f:a6p1n_flareloops} compares flare loops and reconnected
field lines in the simulation. The selected observation times
correspond to $t=2.8\tau_\mathrm{A}$ and $t=7.9\tau_\mathrm{A}$ in the
simulation if the scaling of Section~\ref{ss:scaling} is adopted. A
similar comparison with field lines of the corresponding MFR model is
given in Figure~12 of Paper~I. The two sets of green and red field
lines are traced from start points at $z=0$ in the positive polarity
which lie slightly inside the current layer that extends from the
bottom tip of the vertical current sheet at 2.8 and
$7.9\tau_\mathrm{A}$, so that they are to be compared with the flare
loops in panels~(c) and (d), respectively. (In the figure, both sets
are traced in the field at $7.9\tau_\mathrm{A}$; the lower set is
visually nearly indistinguishable from the field lines traced in the
field at $2.8\tau_\mathrm{A}$.) The trend of progressively decreasing
shear is very clear in the simulation data as well, see the vertical
view in panel~(a). The perspective view at an inclination of
$25^\circ$ from vertical in panel~(b) is chosen to correspond to the
latitude of $\approx25^\circ$ of the active region (the weak tilt
corresponding to the slightly eastward longitude is not taken into
account). Here the field line shapes appear basically similar to the
shapes of the flare loops. The loops possess slightly higher shear.
This is not unexpected, since the use of the potential field in
constructing the NLFFF models is likely to remove some of the shear
that has accumulated in the ambient field in the earlier evolution of
the AR. Part of this shear is recovered in the MFR phase,
and the remaining difference appears to be minor for the considered
region.

A more significant and expected difference is revealed by the
distance of the flare ribbons to the PIL. Compared to the end points
of the newly reconnected field lines at $7.9\tau_\mathrm{A}$, the
separation of the flare ribbons at 03:23~UT is larger by a factor
$\approx1.5$, which signifies a corresponding difference in the
reconnection rate. Given that no attempt was made to correctly model
the reconnection rate (beyond the choice of the numerical diffusion
such that the current sheet does not develop numerical artefacts), the
difference is surprisingly small. We expect that it can be removed by
including resistivity.

The rise profile of the eruption is plotted in
Figure~\ref{f:a6p1n_cfl}. Here the initial, mainly upward-directed
displacement leads to a doubling of the initial apex height (from
$h_0=0.047$ to $h(2)=0.098$) within a time frame similar to the
relaxation runs. The flux rope has roughly semicircular shape at this
time. As soon as the initial velocity enhancement has largely decayed,
one can see a more gradually developing rise with a significant
horizontal component (obvious from the increasing difference between
the solid and dashed curves in the figure). The logarithmic display
reveals that the more gradual rise is approximately exponential (but
it can also be approximated by a power law with an exponent near 3).
The different functional form and direction of the rise after
$t=2\tau_\mathrm{A}$ suggest that the two phases have different
physical origins. This is supported by the evolution of
$\sigma_j(t)$, which first decreases monotonically to
$\sigma_j(t\!=\!3.1)=0.038$ and subsequently increases monotically.
We interpret the initial upward displacement as the
motion to the equilibrium position of the flux rope which it had not
yet reached after the 30,000 MFR iterations that define the initial
condition for the MHD run. The subsequent exponential-to-power law
rise is characteristic of an instability launched by a
small-to-moderate perturbation \citep{Schrijver&al2008b} from the
obviously unstable equilibrium position.

The fluid element at the magnetic axis of the flux rope experiences
the reflection at the side boundary of the box after
$t\approx8\tau_\mathrm{A}$. 8.7 (39) percent of the initial total
(free) magnetic energy are released at this stage. The subsequent part
of the rise profile is not related to reality and included only for
completeness. The additional upward acceleration in the process of the
reflection results from the induction of currents as the expanding
flux is compressed at the side boundary.


\subsection{Scaling the Simulation to the Observations}
\label{ss:scaling}

\begin{figure*}[t]                                              
\centering
\includegraphics[width=.73\linewidth]{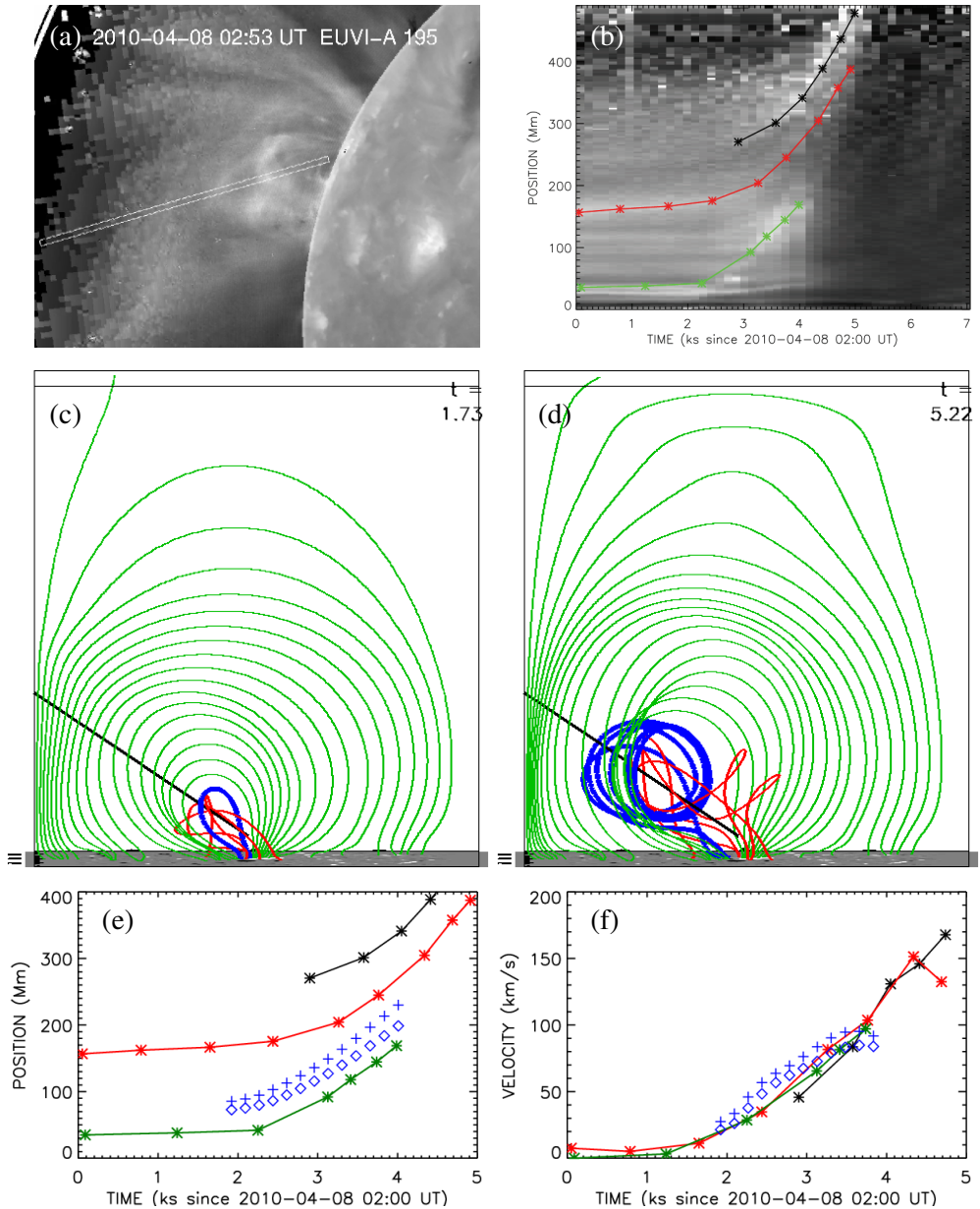}
\caption{\label{f:a6p1n_scaling}
 Scaling of the simulation to the observation data.
 (\emph{a}): A representative EUVI-A 195~{\AA} radial-filter image
 showing the structure of the overlying flux. The indicated artificial
 slit is aligned with the main expansion direction of $45^\circ$
 from vertical and used to generate the stack plot in
 panel~(\emph{b}).
 (\emph{c--d}): Field line plots during the nearly exponential rise
 phase, viewed in $-x$ direction. The field line used for the scaling
 is marked in blue.
 (\emph{e--f}): Scaling of simulation data (blue diamonds and plus
 signs) to the rise profiles marked in the stack plot.
 (A color version of this figure and an animation of the EUVI-A
 195~{\AA} image sequence are available in the online journal.)
}
\end{figure*}

From the rise profile in Figure~\ref{f:a6p1n_cfl} it is obvious that
the simulation can capture properties of the eruption on 2010 April~8
in the early, approximately exponential phase,
$2\lesssim t/\tau_\mathrm{A}<8$. The quantitative comparison between
this part of the rise profile and EUVI-A 195~{\AA} data allows us to
scale the simulation to the observed event. Since the erupting flux
propagates mainly southwards (as seen in the AIA images),
\textsl{STEREO Ahead} observes its rise nearly exactly from the side,
presumably with relatively small projection effects. Note that only
the scaling of the time unit $\tau_\mathrm{A}$, or equivalently of the
velocity unit $\hat{V}_\mathrm{A0}$, is to be determined, since the length
and field strength units are given by the magnetogram and the density
is fixed by field strength and Alfv\'en velocity.

An animation of the EUVI-A 195~{\AA} images is included with
Figure~\ref{f:a6p1n_scaling}, and panel~(a) shows a representative
image. No trace of the rising filament material can be seen, but
several overlying loops stand out relatively clearly. Their rise is
quantified in the stack plot (panel~(b)) taken at an artificial slit
which is aligned with the average direction of ascent (see panel~(a)).
The three clearest traces are marked in color.

Next, we select field lines which correspond to the overlying loops.
It turns out that no exact match is possible, since the usable height
range in the simulation falls between the lower two traces in the
stack plot. This is seen in Figures~\ref{f:a6p1n_scaling}(c) and (d), where
a ray is indicated, corresponding to the artificial slit in the
EUVI-A image. The panels display a view in $-x$ direction, similar to
the view of \textsl{STEREO Ahead}. The ray starts at the apex of the
flux rope's magnetic axis at $t=0$ and lies in the $y$-$z$ plane,
inclined to the vertical by $45^\circ$. After the initial upward
displacement, the flux rope radius has grown such that the innermost
overlying flux lies at a distance of about 70~Mm along the ray, larger
than the pre-eruption distance of $\approx40$~Mm of the green trace in
the stack plot. The pre-eruption distance of the red trace is about
175~Mm, already relatively close to the side boundary of the box. Flux
in this height range can be followed until the compression zone near
the side boundary is encountered, roughly at 250~Mm along the ray, but
this is, of course, stronger influenced by the side boundary than the
flux immediately surrounding the rope, and it can only utilize a small
part of the red trace in the stack plot. Nevertheless, using a field
line in the innermost overlying flux permits an acceptable scaling,
since the expansion of the overlying flux is rather uniform: the three
traces run approximately parallel to each other, i.e., with similar
velocity profiles.

We thus select a fluid element on the ray which lies slightly outside
the flux rope at $t\approx1.75\tau_\mathrm{A}$, the beginning of
nearly exponential rise. The trajectory of this fluid element is
computed through the simulation. At later times, a field line is
traced from the new position and projected onto the $y$-$z$ plane,
where its crossing with the ray yields a new projected distance along
the ray. Finite differencing yields a projected velocity along the
main direction of propagation for both observation and simulation
data.

The EUVI-A images (of 2.5~min cadence) place the onset time of the
expansion in a relatively narrow interval, 02:30:30--02:33:00~UT, of
which we associate a time near the midpoint, 02:32~UT, with
$t=1.75\tau_\mathrm{A}$ in the simulation. The value
$\hat{V}_\mathrm{A0}\approx1400$~km\,s$^{-1}$ that yields a good match of
both the projected distance and velocity data is then easily found by
trial and error; see the diamonds in
Figures~\ref{f:a6p1n_scaling}(e) and (f), which cover the interval
$t=(1.73\mbox{--}5.93)\tau_\mathrm{A}$ in the simulation. A change of
$\pm100$~km\,s$^{-1}$ from the estimated $\hat{V}_\mathrm{A0}$ already
produces a much poorer match.

We note that a trend of decreasing acceleration becomes visible in the
projected velocity data of the overlying flux in the simulation
already in the middle of the considered interval. This is not seen in
the observation data and indicates how early the overlying flux is
influenced by the side boundary. Nevertheless, the match between
simulation and observation appears acceptable up to the second-to-last
velocity point in the interval.

An ambiguity in the above procedure originates from the insufficient
knowledge about the position of the overlying loops relative to the
plane of sky for \textsl{STEREO Ahead}. They cannot be unambiguously
identified in the corresponding AIA 193~{\AA} images. We know from AIA
that the main direction of expansion was nearly southward, i.e.,
nearly in the plane of sky for \textsl{STEREO Ahead}, but that the
initial expansion had a strong southeastward component. To roughly
estimate how this ambiguity might influence the scaling, we repeat the
procedure, positioning the fluid element at
$t\approx1.75\tau_\mathrm{A}$ on rays directed $\pm30^\circ$ off the
meridional plane (but also starting at the flux rope apex). For the
southeastward pointing ray, the same $\hat{V}_\mathrm{A0}$ is obtained
from a match of nearly the same quality (see the plus signs in
Figures~\ref{f:a6p1n_scaling}(e) and (f)). For the southwestward pointing
ray, a higher Alfv\'en velocity is indicated (by $\sim50\%$), but it
is not possible to match both distances and velocities to the stack
plot data at a reasonable quality. The AIA data do not show any
distinctive structure propagating in this direction in the time
interval used in the scaling. Thus, we consider it very unlikely that
the structures seen by EUVI-A have expanded in this direction. On the
other hand, the consistency of the $\hat{V}_\mathrm{A0}$ estimates, based on
the southward and southeastward directed rays, makes them rather
reliable. For completeness we note that the considered fluid elements
remain relatively close to their respective ray in the course of the
eruption.

The estimated value of the peak initial Alfv\'en velocity
$\hat{V}_\mathrm{A0}$ refers to the point of peak field strength in the box,
$\max(B_0)=955$~Gauss, which lies in the magnetogram plane. Our model
for the initial density implies
$V_\mathrm{A0}\propto{B_0^{1/4}}$, so that the initial field
strength of $\approx70$~Gauss in the area of the flux rope apex yields
an initial Alfv\'en velocity $V_\mathrm{A0}\approx730$~km\,s$^{-1}$
and initial particle density $N_0\approx4.4\times10^{10}$~cm$^{-3}$ in
the upper part of the flux rope. The density must be taken as a
characteristic average density in the filament channel, since our
density model disregards any filamentary fine structure of the plasma.

One has to consider this velocity as the lower end point of a
plausible range (and the density, correspondingly, as an upper end
point). This value follows from a model with a certain amount of
prescribed axial flux, $\Pa=6\times10^{20}$~Mx, which lies near the
point of marginal stability, but at a somewhat arbitrary distance set
by the interval of $10^{20}$~Mx between the $\Pa$ values in
consecutive models. An unstable model with less axial flux would yield
a slower rise in the simulation, consequently, a higher estimate for
$\hat{V}_\mathrm{A0}$. Indeed, the leading-edge CME velocity of
$\approx520$~km~s$^{-1}$ observed by COR2 is at 70\% of the Alfv\'en
velocity in the source. Allowing for a factor 2 lower velocity of the
CME core, this becomes ${\sim}\,V_\mathrm{A0}/3$, still a
rather high value for a moderate CME that gains its acceleration in a
large height range of several $R_\odot$. Such eruptions tend to reach
lower terminal flux rope velocities of
$\lesssim0.2V_\mathrm{A0}$ \citep{Torok&Kliem2007}. Thus, a
range for the Alfv\'en velocity in the flux rope from
$\approx730$~km\,s$^{-1}$ up to about a factor 2 higher appears
reasonable. The corresponding density range is
$\sim(1\mbox{--}4)\times10^{10}$~cm$^{-3}$.

These estimates lie well within acceptable ranges. \citet{Gary2001}
finds $\beta\sim0.02\mbox{--}0.2$ at a height of $0.1~R_\odot$ (see
his Figure~3), which translates to
$V_\mathrm{A}=2C_\mathrm{s}\beta^{-1/2}\sim(650\mbox{--}2000)$~km~s$^{-1}$
for a plasma temperature of $T=2.5$~MK (i.e., a sound speed of
$C_\mathrm{s}=145$~km~s$^{-1}$). The lower part of the range is
appropriate for a considerably dispersed, moderate AR like
AR~11060. \citet[][their Section~3.3.2]{Labrosse&al2010} quote the
wide range of $N\sim(10^9\mbox{--}10^{11})$~cm$^{-3}$ for densities of
cool (H$\alpha$-emitting) to moderately hot (EUV-emitting) plasmas in
quiescent and AR prominences. The lower-to-middle part of
the range may be representative for the average density in the
erupting flux in the considerably dispersed AR.


\subsection{Onset Condition}
\label{ss:criterion}

By reaching good agreement with key features of the eruptive event on
2010 April~8, like the initial height-time profile and the rise
direction of the erupting flux, our simulations substantiate the
conclusion in Paper~I that the flux rope insertion method allows to
model the NLFFF in AR~11060 around the time of the eruption. The
condition for the loss of equilibrium, formulated as a condition of
flux imbalance in terms of the ratio between the rope's axial flux and
the unsigned flux in the source region of the eruption, is also
confirmed. Given the total unsigned flux in AR~11060 of
$F_\mathrm{u}=3.7\times10^{21}$~Mx (Paper~I), the range of
$\Pa=(5\mbox{--}6)\times10^{20}$~Mx for the marginal stability point
yields a flux ratio $\Pa/(F_\mathrm{u}/2)=(27\mbox{--}32)\%$, well
within the range found previously \citep{Bobra&al2008,
Savcheva&al2012c}. A similar flux ratio is reached at the onset of
eruption in the simulation that applied flux cancellation to the model
with $\Pa=5\times10^{20}$~Mx.

It is of principal interest to compare this condition with the
condition obtained by describing the loss of equilibrium as a plasma
instability, i.e., the kink instability of a current channel. The
relevant mode of the kink instability here is the lateral kink, known
as the torus instability in the case of an arched current channel,
since the pre-eruptive NLFFF possesses insufficient twist for the
excitation of the helical kink mode. Using the length and flux values
of the two flux rope models next to the marginally stable point,
$L\approx270$~Mm, $\Pa=(5\mbox{--}6)\times10^{20}$~Mx, and
$\Fp=10^{10}$~Mx~cm$^{-1}$, the average twist angle is
$LB_\phi/(rB_z)\sim{\pi}L\Fp/\Pa=(0.53\mbox{--}0.45)\pi$, far smaller
than the threshold of the helical kink.

The threshold of the torus instability is given in terms of the decay
index of the external poloidal field, $B_\mathrm{ep}$, at the position
of the current channel,
$n=-d\ln{B_\mathrm{ep}}/d\ln{z}>n_\mathrm{cr}$. The canonical values
of the critical decay index are $n_\mathrm{cr}=3/2$ for a toroidal
current channel \citep{Bateman1978} and $n_\mathrm{cr}=1$ for a
straight current channel \citep{vanTend&Kuperus1978}.
\citet{Kliem&Torok2006} showed that the value for a toroidal current
channel rises to $n_\mathrm{cr}=2$ if reconnection under the flux rope
does not commence immediately, and they added a relatively small
negative correction term. \citet{Demoulin&Aulanier2010} generalized
the consideration to arbitrarily shaped paths of the current channel,
finding a range $n_\mathrm{cr}\approx1.1\mbox{--}1.3$. Most numerical
studies, including the only parametric study to date, find
$n_\mathrm{cr}$ in the range 1.5--1.75 \citep{Torok&Kliem2007,
Aulanier&al2010, Fan2010}, but it should be noted that the simulation
in \citet{Fan&Gibson2007} yields a value of 1.9. The range 1.5--1.75
is also supported by observational studies \citep{Y.Liu2008,
Y.Guo&al2010}.  The critical decay index must depend upon the
photospheric line tying (which varies with the shape of the current
channel) and it should increase for increasing strength of the
external toroidal (shear) field. These dependencies have not yet been
investigated. However, we can expect these effects to be relatively
minor in the present case, since the line tying is minimized for
semicircular flux rope shape, which the unstable flux rope reaches at
the onset of the eruption, and since the ambient potential field
has only little shear. In the comparison below, we adopt the currently
best supported range of the torus instability threshold,
$n_\mathrm{cr}\approx1.5\mbox{--}1.75$.

\begin{figure}[t]                                               
\centering
\includegraphics[width=.875\linewidth]{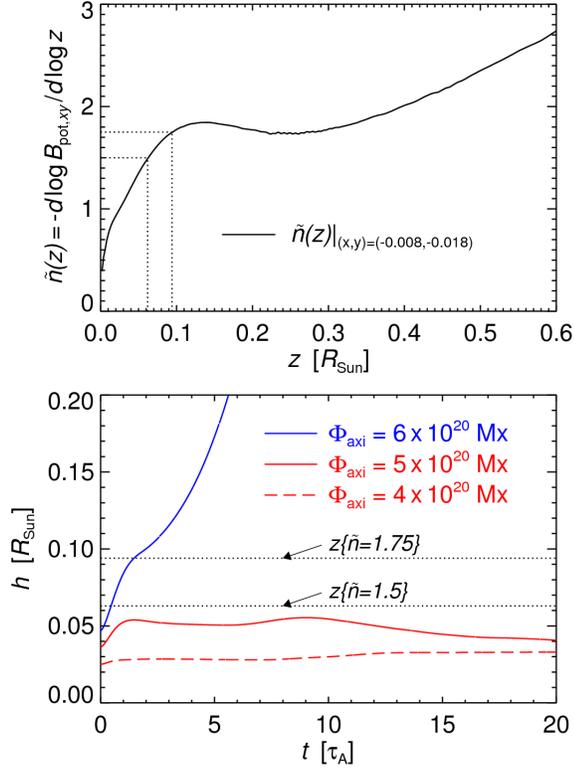}
\caption{\label{f:criterion}
 (a) Decay index $\tilde{n}=-d\ln{B_\mathrm{pot,\,\textit{xy}}}/d\ln{h}$
 of the horizontal components of the potential field along a vertical
 line passing through the apex point of the flux rope's magnetic axis
 at $t=0$.
 (b) Height-time profiles of the flux rope apex for the three models
 studied in this paper, compared to the onset condition of the torus
 instability.
 (A color version of this figure is available in the online journal.)
}
\end{figure}

In practice it is often impossible or rather laborious to isolate the
external poloidal field $B_\mathrm{ep}$. This involves the subtraction
of the poloidal field created by the current channel in the corona,
which requires the exact knowledge of the current channel, i.e.\ of
the NLFFF. Here we have this knowledge from the relaxed configuration
with $\Pa=5\times10^{20}$~Mx and from the initial configuration with
$\Pa=6\times10^{20}$~Mx. However, given the effort required for this
analysis and given the uncertainty in the knowledge of the critical
decay index, we choose to simplify the consideration by instead using
the decay index of the horizontal component of the (total) potential
field, $\tilde{n}=-d\ln{B_\mathrm{pot,\,\textit{xy}}}/d\ln{z}$. We
have verified that the potential field passes at nearly right angles
over the PIL in the height range of the flux rope at $t=0$, so that it
is a reasonable approximation of the external poloidal field. The
direction is much closer to perpendicular than that of the immediately
overlying flux in the initial configuration of the MHD simulation
(green field lines in Figure~\ref{f:a6p1n_FR+overl} at $t=0$), which
is considerably influenced by the presence of the flux rope in the
course of the MFR. We have computed $\tilde{n}(z)$ at several
positions along the PIL and found it to be nearly uniform in the
section of the PIL between the main flux concentrations where the
middle part of the inserted flux rope is located.
Figure~\ref{f:criterion} shows the profile at the vertical line
through the flux rope apex. The height range corresponding to the
adopted range of $n_\mathrm{cr}$ is $z=0.063\mbox{--}0.094$.

The bottom panel of Figure~\ref{f:criterion} summarizes the
height-time profiles of the three models studied in this paper. The
apex of two stable ones stays in the range
$h\approx0.03\mbox{--}0.055$, i.e.\ below the torus-unstable height
range. The inferred equilibrium apex height of the unstable model,
$h(t\!=\!2)=0.098$, falls rather clearly in the torus-unstable height
range (as discussed above, the rise up to the equilibrium height
is not related to an instability).
Thus, for the considered AR~11060, the stability boundary given
in terms of the axial flux in the rope,
$\Pa=(5\mbox{--}6)\times10^{20}$~Mx, corresponds very well to the most
likely range for the threshold of the torus instability, 
$n_\mathrm{cr}\approx1.5\mbox{--}1.75$.


\section{Summary and Conclusions}
\label{s:conclusions}

This paper reports MHD simulations that study NLFFF models of the
eruptive AR~11060. The models were constructed in Paper~I
\citep{Su&al2011} using the flux rope insertion method and
MFR. We selected three that appeared to best
represent the pre-eruptive, nearly marginally stable, and eruptive
states of the active region, differing only in the axial flux of the
inserted flux rope, which encompasses the range
$\Pa=(4\mbox{--}6)\times10^{20}$~Mx.
Our simulations confirm key results of Paper~I:

\begin{enumerate}
\item 
There is a limiting value of the axial flux in the rope for
the existence of stable NLFFF equilibria, which lies in the selected
range. The equilibrium height of the flux rope increases with axial
flux.

\item 
Stable models relax deeply in the MHD simulation, reaching
numerical equilibria that retain a flux rope in the considered range
of $\Pa$. Field lines in the flux rope and its immediate surrounding
correspond well in shape to the observed coronal loops.

\item 
The flux rope in the unstable model experiences a full
eruption.
\end{enumerate}

Thus, although none of the observed structures in AR~11060 shows
regular multiple crossings, the definite signature of a flux rope, our
simulations substantiate the conclusions in Paper~I that a flux rope
existed in the active region prior to the eruption and that the rope's
loss of equilibrium caused the eruption.
Additionally, we obtain the following results:

\begin{enumerate}
\setcounter{enumi}{3}
\item 
The model with $\Pa=5\times10^{20}$~Mx, found to be nearly
marginally stable in Paper~I, is also found to be closest to marginal
stability---on the stable side. It does not erupt even though the HFT
at its underside pinches into a short vertical current sheet and a
transient phase of reconnection in this current sheet is triggered,
lasting for a couple of Alfv\'en times. The model can be driven to
eruption by photospheric flux cancellation, and the ratio of its axial
and unsigned flux at that point is similar to the flux ratio of the
unstable model.

\item 
The erupting flux rope shows an accelerated rise $h(t)$,
which can be approximated by an exponential and also by a power-law
with an index near 3. Such a rise is characteristic of a flux rope
instability \citep{Schrijver&al2008b}. The rise velocity reaches a
considerable fraction of the initial Alfv\'en velocity in the flux
rope, about 20\% (still rising when the boundary of the simulation box
is approached). This indicates that the model lies clearly in the
unstable domain of parameter space.

\item 
The simulated eruption agrees very well with several
characteristics of the modeled event, including
(\textit{i}) the initially very inclined direction of ascent, by
$45^\circ$ from vertical toward the equator;
(\textit{ii}) the accelerated rise of the overlying flux; and
(\textit{iii}) the close temporal association between the acceleration
of the flux rope (the CME component of the eruption) and the
development of reconnection in the vertical current sheet underneath
(the flare component).
Additionally, initially high shear and the trend of progressively
decreasing shear in the reconnected flux under the rope, as displayed
by the flare loops, are clearly reproduced. However, there are also
differences.
(\textit{i}) The simulated rise begins to turn more radial in the low
corona, whereas the modeled event showed an increasing
inclination in the low corona, turning more radial only at heights of
several solar radii.
(\textit{ii}) The footpoint locations of newly reconnected field lines
in the simulation move away from the PIL at a somewhat lower speed
than the observed flare ribbons, indicating a moderately lower
reconnection rate. Obvious reasons and straightforward options for
improvement of the modeling were identified for both.

\item 
Oblique eruption paths can be caused by the specifics of the
magnetic structure in the source region, in addition to the
interaction with a coronal hole and the asymmetry of the photospheric
flux distribution suggested earlier \citep{Gopalswamy&al2009,
Panasenco&al2013}. We suggest that the slingshot action of
horizontally curved flux, which was, or quickly became, part of the
erupting flux, contributed to the inclination of the rise path in the
modeled event.

\item 
The ability to follow the eruption allows us to scale the
simulation to the observation data, thus estimating the Alfv\'en
velocity in the source region of the eruption. The range
$V_A\sim(730\mbox{--}1500)$~km\,s$^{-1}$ in the volume of the flux
rope is indicated. Since the field strength in the rope is given by
the NLFFF model, $B\approx70$~Gauss, the estimated Alfv\'en velocity
implies an average flux rope density in the range
$N\sim(1\mbox{--}4)\times10^{10}$~cm$^{-3}$. Both values lie within
acceptable ranges, especially the Alfv\'en velocity, since AR~11060
was already considerably dispersed.

\item 
By its sharper discrimination between stability and
instability compared to the magnetofrictional relaxation technique,
the MHD modeling yields a narrower range for the threshold value of
the considered control parameter. We find that the value lies in the
range $\Pa=(5\mbox{--}6)\times10^{20}$~Mx (compared to
$(4\mbox{--}6)\times10^{20}$~Mx in Paper~I). This corresponds to a
range of the flux ratio $\Pa/(F_\mathrm{u}/2)=0.27\mbox{--}0.32$,
where $F_\mathrm{u}=3.7\times10^{21}$~Mx is the total unsigned flux in
the AR.

\item 
This range of the flux ratio corresponds very well to the
threshold of the torus instability in the considered active region.
Using the horizontal component of the potential field as a proxy for
the external poloidal field component, the decay index
$n=-d\ln{B_\mathrm{ep}}/d\ln{z}$ takes values
$n\approx1.3\mbox{--}1.8$ in the range of equilibrium flux rope
heights found for the range of axial flux
$\Pa=(5\mbox{--}6)\times10^{20}$~Mx. The critical decay index for
onset of the torus instability is estimated to lie in the range 
$n_\mathrm{cr}\approx1.5\mbox{--}1.75$. This establishes a connection
between these independently developed criteria for the loss of
equilibrium of a flux rope, which is plausible from the following
consideration. Higher axial flux in the rope corresponds to greater
equilibrium height (as a consequence of the implied higher flux rope
current), which, in turn, typically corresponds to higher decay index,
i.e., to approaching the torus-unstable height range where
$n>n_\mathrm{cr}$.

\item 
A considerable amount of changes were found while the stable
models relaxed to a numerical equilibrium in the MHD simulation,
although they had already gone through an MFR
phase. The MFR had a pre-set number of
iterations, thus it had not progressed to the deepest level possible,
especially not for the nearly marginally stable model, but the model
with  $\Pa=4\times10^{20}$~Mx was quite well relaxed. Even changes of
the topology (a transient split of the flux rope) occurred in the MHD
relaxation. Only a part of the different level of dynamics appears to be
due to the numerical differences between the employed
magnetofrictional and MHD simulation codes. A systematic comparison of
magnetofrictional vs.\ MHD relaxation might thus be warranted.

\end{enumerate}

Based on various developments and applications of the flux rope
insertion method \cite[e.g.,][]{vanBallegooijen2004, Bobra&al2008,
Su&al2011}, this research represents a further step toward
data-constrained numerical modeling of solar eruptions. Future work
along this line will consider coronal NLFFF constructed from a time
series of magnetograms, as in \citet{Savcheva&vanBallegooijen2009} and
\citet{Savcheva&al2012c}, and eventually proceed to driving coronal
NLFFF through time-dependent photospheric boundary data obtained from
observations. Technical improvements toward using larger domains and
open boundary conditions will permit following eruptions much further
than possible here; this will set tighter constrains as well.

\begin{acknowledgements}

We thank the referee for a careful reading which was helpful
in improving the clarity of the presentation.
We acknowledge the use of data provided by the \textsl{Hinode}/XRT,
\textsl{STEREO}/SECCHI, and \textsl{SDO}/AIA and HMI instruments.
This paper also uses data from the CACTus CME catalog, generated and
maintained by the SIDC at the Royal Observatory of Belgium. 
BK acknowledges support by the DFG, the STFC, and the NSF (Grant
AGS-1249270). He also acknowledges
the hospitality of the solar group at the Yunnan Astronomical
Observatory, where part of this work was completed, and the associated
support by the Chinese Academy of Sciences under grant no.~2012T1J0017.
This project is partially supported under contract NNM07AB07C from
NASA to the Smithsonian Astrophysical Observatory (SAO) and SP02H1701R
from LMSAL to SAO.

\end{acknowledgements}


\bibliographystyle{apj}
\bibliography{ms8apr}

\end{document}